\documentclass[12pt]{iopart}% IOP page structure
\usepackage{url}% writing URLs
\usepackage{amsmath}% equations formatting
\usepackage{amssymb}
\usepackage{siunitx}% SI units
\usepackage{booktabs}% table formatting
\usepackage{makecell}% more control over table cells
\usepackage{tabularx}% more control over table dimensions
\usepackage{graphicx}% figures formatting
\usepackage{rotating}% allowing things to rotate
\usepackage{pdfpages}% importing PDFs
\usepackage{pdflscape}% importing PDFs in landscape
\usepackage[colorinlistoftodos,textsize=tiny,textwidth=0.9in]{todonotes} % for margin comments in pdf
\usepackage{marginnote}
\newcommand{\todotoo}[2][]{{% manual command for left margin comments in pdf
 \let\marginpar\marginnote
 \reversemarginpar
 \renewcommand{\baselinestretch}{1.0}%
 \todo[#1]{#2}}}

\usepackage{multirow}

\begin{document}

\title[Training and Upgrading Tokamaks]{Training and Upgrading Tokamak Power Plants with Remountable Superconducting Magnets}

\author{S. B. L. Chislett-McDonald$^1$, E. Surrey$^2$, J. Naish$^2$, A. Turner$^2$ and D. P. Hampshire$^1$}

\address{$^1$Superconductivity Group, Centre for Materials Physics, Department of Physics, Durham University, UK}

\address{$^2$Culham Centre for Fusion Energy, Culham Science Centre, Abingdon, UK}

\ead{simon.chislett-mcdonald@durham.ac.uk}

\begin{abstract}

All high field superconductors producing magnetic fields above 12 \si{\tesla} are brittle. Nevertheless, they will probably be the materials of choice in commercial tokamaks because the fusion power density in a tokamak scales as the fourth power of magnetic field. Here we propose using robust, ductile superconductors during the reactor commissioning phase in order to avoid brittle magnet failure while operational safety margins are being established. Here we use the \texttt{PROCESS} systems code to inform development strategy and to provide detailed capital-cost-minimised tokamak power plant designs. We propose building a `demonstrator' tokamak with an electric power output of 100 \si{\mega\watt}$_\text{e}$, a plasma fusion gain $Q_\text{plasma}$ = 17, a net gain $Q_\text{net}$ = 1.3, a cost of electricity (COE) of \$ 1148 (2021 US) per \si{\mega\watt\hour} (at 75 \% availability) and high temperature superconducting operational TF magnets producing 5.4 \si{\tesla} on-axis and 12.5 \si{\tesla} peak-field. It uses Nb-Ti training magnets and will cost about \$ 9.75 Bn (2021 US). An equivalent 500 \si{\mega\watt}$_\text{e}$ plant has a COE of \$ 608 per \si{\mega\watt} suggesting that large tokamaks may eventually dominate the commercial market. We consider a range of designs optimised for capital cost (as the reactors considered are pilot plants) consisting of both 100 \si{\mega\watt}$_\text{e}$ and 500 \si{\mega\watt}$_\text{e}$ plants with each of two approaches for the magnets: training and upgrading. With training magnets, the  plant is cost-optimised for REBCO TF magnets. For a 100 \si{\mega\watt}$_\text{e}$ plant, the Nb-Ti training magnets typically produce 70 \% peak field on the toroidal field coils compared to REBCO magnets, 65 \% peak field on the central solenoid and cost $\approx$ 10 \% of the total machine cost. Training magnets could in principle be reused for each of say 10 subsequent (commercial) machines and hence at 1 \% bring only marginal additional cost. With upgrade magnets the plant is more expensive - first it is cost-optimised for Nb-Ti and then upgraded to REBCO coils. The upgrade increases the net electrical output from 100 to 280 \si{\mega\watt}$_\text{e}$ with an $\approx$ 25 \% increase in reactor capital cost. We also evaluate likely advances in fusion technology and find that technologies on the horizon will probably not bring further large reductions in capital cost, and that REBCO magnets are generally stress-limited rather than current density limited. We conclude that: the fusion community should develop high $B_{c2}$ alloys specifically for fusion applications; superconductors should be tested under operational-like radiation at cryogenic temperatures; and that we should proceed now with detailed design and construction of a prototype fusion power plant that integrates and de-risks all the key technologies including high temperature superconducting cables and joints using remountable training magnets, and hence is the last tokamak before commercialisation of fusion energy. 

\end{abstract}


\section{Introduction} \label{Introduction}
High temperature superconducting REBCO (Rare-Earth barium copper oxide) materials and the low temperature superconductor Nb$_3$Sn are the candidate high field materials for the toroidal field (TF) and central solenoid (CS) coils for fusion reactors. The ITER \cite{aymar02} and SPARC \cite{creely20} reactors use these materials and are expected to operate with fusion plasma gain $Q_\text{plasma} \approx 10$, as will pilot fusion power plants that will eventually generate 100s \si{\mega\watt} net electricity (\si{\mega\watt}$_\text{e}$) (e.g. EU-DEMO \cite{federici18}, STEP \cite{STEP20}, ARC \cite{sorbom15}). During the commissioning phase after construction, in addition to the high stresses that occur in magnets during standard operation, unexpected powerful disruptions can also occur in the plasma such as vertical displacement events and associated halo currents.  
In the JET reactor, such uncontrolled events induce forces of order 4 \si{\mega\newton}, that have lifted the entire vessel by 9 \si{\milli\metre} \cite{Balshaw18,riccardo04}. These disruptions are expected to be an order of magnitude greater in ITER \cite{riccardo10}. More than half of all unintentional disruptions in JET were not due to physics instabilities, but were attributed to, for example, failure of one of the sub-systems, control errors or human error. Fewer disruptions predominantly followed better technical operation of JET \cite{devries11}. In short, operating a tokamak properly requires completion of a `learning curve' that does not require full plasma power operation \cite{devries11}. These uncontrolled events would bring with them the risk of permanent and irreparable damage to a tokamak including the expensive brittle superconducting magnets. In ITER, were a TF coil failure to occur before nuclear operation starts, one can reasonably expect it to take 4 years to replace the coil (using an available spare) \cite{private_communication}. Once nuclear (i.e. D-T) operation has begun, the activation in the reactor vessel would be so high that, given there is no robotic control of magnet replacement, it would probably not be cost-effective to replace a TF coil \cite{gilbert18}. Unfortunately all the very high field superconductors that we need in full operation for optimal (profitable) commercial fusion are brittle, and the largest Nb$_3$Sn fusion magnets ever produced to-date (by size or weight) were resistive when first turned on \cite{lubell88}. In this paper, we propose using Nb-Ti during the critical commissioning and testing phase because although it has poorer high field performance, it is ductile and so is robust against mechanical or brittle failure.

There are a number of road-maps \cite{federici18,NIST21,tobita19} that focus on achieving a final build machine. It is now understood that both climate change and commercial imperatives mean that one simply can't wait 15 years to further optimise plasma performance \cite{mitchell21}. Indeed U.N. Secretary-General António Guterres described the recent UN IPCC (United Nations Intergovernmental Panel for Climate Change) report as a ``code red" for humanity \cite{UN_code_red}. Here, we recommend starting with lower risk machines with remountable magnets that eventually have components that are exchanged for higher-risk higher-performance components only when they are needed \cite{Chislett-McDonald20}. We calculate the cost of building various tokamak designs using the \texttt{PROCESS} systems code \cite{kovari14,kovari16,Process_2022} and specify a commissioning roll-out that helps avoid single-point failure (and while high temperature superconducting cables are improved and made cheaper). We use an approach that takes advantage of new advances (e.g. if new higher field ductile superconductors are developed), so that the expensive and time-consuming tokamak recommended for construction still correctly identifies and de-risks the best available commercial solution. More specifically, in this paper we have considered swapping superconductors using two approaches: training and upgrading and have assumed that all magnets will be fully remountable. With training magnets, the plant is cost-optimised for full power operation with REBCO but trained first using commercial Nb-Ti TF and CS coils (for all tokamaks in this work with REBCO, Nb$_3$Sn and commercial Nb-Ti TF and CS coils, commercial Nb-Ti is used for the PF coils; for all tokamaks in this work with  quaternary Nb-Ti TF and CS coils, quaternary Nb-Ti is used for the PF coils). With the (more expensive) upgrading magnets approach, the plant is cost-optimised for Nb-Ti and then upgraded to REBCO coils. We have used \texttt{PROCESS} to find optimal designs, defined in all cases to minimise plant capital cost. \texttt{PROCESS} reports costs in 1990 US \$, to convert to 2021 costs we have used CPI inflation \cite {CPI_inflation,oishi12,IAEA11} resulting in a conversion factor of 1 US \$ 1990 = 2.13 \$ in 2021. An alternative is the IHS-CERA index for nuclear fission power plant costs \cite{IHS_CERA}. The index data were collected between 2000 and 2017, we therefore use the consumer price index (CPI) to extrapolate from the IHS-CERA index to 1990 and 2021. Using this metric, 1 US \$ 1990 = 3.28 \$ in 2021 (when spent in the nuclear power sector). One must therefore take care when converting to today's costs as they can vary significantly depending on the choice of index.

For each of the reactor training and upgrading approaches, we have considered three power plant designs which gives us six baseline tokamaks. In each approach two tokamaks produce 100 \si{\mega\watt}$_\text{e}$ and are designed for H$_\text{98}$ = 1.2 and 1.6 (to encompass future advances) and the third produces 500 \si{\mega\watt}$_\text{e}$ and is designed for H$_\text{98}$ = 1.2. For each of these six baseline designs we have then investigated swapping superconductors out whilst maintaining the baseline reactors' architectures. Table \ref{ReactorComparison} shows the key design and performance parameters for the six baseline tokamaks (including our preferred choice for build - a 100 \si{\mega\watt}$_\text{e}$, H$_\text{98}$ = 1.2, pilot power plant optimised for REBCO), together with the most important tokamaks that operated, are operating or are planned. Here we use the term 'demonstrator' for those tokamaks that are intended to be a last tokamak before commercialisation, and therefore de-risk all the key technologies. Important issues concerning remote handling and remountable coils are considered in section \ref{Remountable coils}. Section \ref{Design choices} details the reactor design choices associated with considerations of  plasma physics and engineering design. The design of optimised radiation shield used for all reactor designs is described in section \ref{MCNP shield} using detailed \texttt{MCNP} \cite{goorley12} calculations.  In section \ref{Superconductors} we describe the cost models for current superconductors and the impact of future developments. Our preferred reactor design, the other five capital-cost-minimised reactor designs and the performance of all swapped superconducting magnets for each design are presented in section \ref{Results}. Future developments of decreased REBCO cost and increased steel yield stresses on our preferred reactor design (as a proxy for both improved materials and better magnet design) are discussed in section \ref{Future developments}. The range of tokamaks and preferred choice is discussed in section \ref{Discussion}. Finally we summarise the most important results in section \ref{Conclusions}.

\section{Robotics, Remountable Magnets and Joints}\label{Remountable coils}

In high aspect ratio reactors of the type considered here (similar to ITER and EU-DEMO), robotics/crane systems can in principle be employed to extract the remountable CS and TF coils if they become damaged because they see little neutron flux due to thick radiation shielding (for example most EU-DEMO coils will be considered Non-Active-Waste at end of life \cite{gilbert17,gilbert18,gilbert19}). The requirements on remote handling (RH) for coils alone, are therefore not particularly demanding in the designs considered in this paper. However, maintenance of the first wall, blanket and the other internal reactor components will be extremely challenging due to the high radiation levels of order 100 - 600 \si{\giga\becquerel\per\kilogram} remaining 4 weeks after shutdown \cite{gilbert17}, similar to those found in the core container of a fission reactor 8 years after shutdown \cite{decomission}. Specialised radiation hardened RH systems as found in EU-DEMO's internal RH system \cite{damiani18,keep17} will be required and have not been included in the costs (for replacing a damaged irradiated magnet) in this paper.  The economic damage of a damaged or destroyed non-remoutable magnet would be unjustifiable: it would put a power plant out of action for years. This risk is simply unacceptable for a commercial plant, and is the reason why (in the author's opinion) ``life-time component'' magnets will not be permitted in future reactors.
Commercial reactors will require availabilities as large as possible (70\% or higher). Magnet failure forces the reactor to shut down, and repairs and replacements must be made as swiftly as possible. Remountable magnets and joints will be required to enable magnet replacement without having to cut open the vacuum vessel and the shielding (using for example a `half-phi' design \cite{hampshire18}). Only the coils themselves would have to be taken apart (with careful engineering and masterful crane/robotics operation). This is as opposed to existing machines in which an entire reactor section must be removed (including components internal to the magnets) for the magnets themselves to then be removed. Remountable joints for both low temperature superconductors \cite{powell80} and high temperature superconductors \cite{sorbom15,hashizume14,hartwig12,mangiarotti15} have been designed, though to-date none have been incorporated into working tokamaks. Indeed there is no published work (at time of writing) demonstrating full-scale remountable joints. There have been preliminary works on cable-to-cable remountable joints \cite{hartwig20,Weiss21} and there are no reasons why such joints couldn't simply be scaled to a full winding pack \cite{bruzzone18a}. With concentrated effort however, we are confident that full-scale, reactor ready remountable joints are feasible by 2035-2040: non-remountable joints in superconducting magnets are commonplace in large-scale magnets; multiple institutions are working together and competitively to build them \cite{hartwig20,bruzzone18a,Weiss21}; remountable joints are already present in resistive magnet tokamaks e.g. MAST-U. Remountable joints introduce additional thermal load on the reactor cryo-system, though the required wall plug cryoplant power to manage this is $\approx$ 1 \si{\mega\watt} \cite{sorbom15} (though, of course, dependent on the specifics of the magnet system) and has been omitted from the \texttt{PROCESS} calculations here as it is small compared to the pre-existing power demand for cooling of order 40 - 50 \si{\mega\watt_e}. For example, soldered REBCO joints have resistances of $\approx$ 50 \si{\nano\ohm\per\square\centi\metre} \cite{tsui16}, which for an ITER TF coil system gives a total thermal load of $\approx$ 600 \si{\watt} and a power demand for additional cooling that is only about 1 \% of the total.

In the current final development phase for commercial fusion, one would not want to use brittle REBCO magnets in the commissioning phase for the reactor when the risk of damaging the magnets is not well-known. Indeed, operating with Nb-Ti training magnets may be required as part of regulatory licensing of the construction prior to operation \cite{taylor17}. After the reactor has operated successfully for several years and completed all its commercial requirements, in the post-demonstrator research reactor phase one may reuse the Nb-Ti magnets as part of trialling new technologies and component designs, because the risk of disruptions during such trials may again be high. 

\section{Reactor Design Choices}\label{Design choices}
In this section, we consider the most important technological areas of tokamak development. We describe the choices and constraints that affect the design and cost-minimisation we have made in each area, explaining the reasoning for our choices.

\subsection{Plasma Operation}

\subsubsection{Confinement Time and H$_\text{98}$-factor}
\begin{figure}
    \centerline{
    \includegraphics[height=9cm,keepaspectratio]{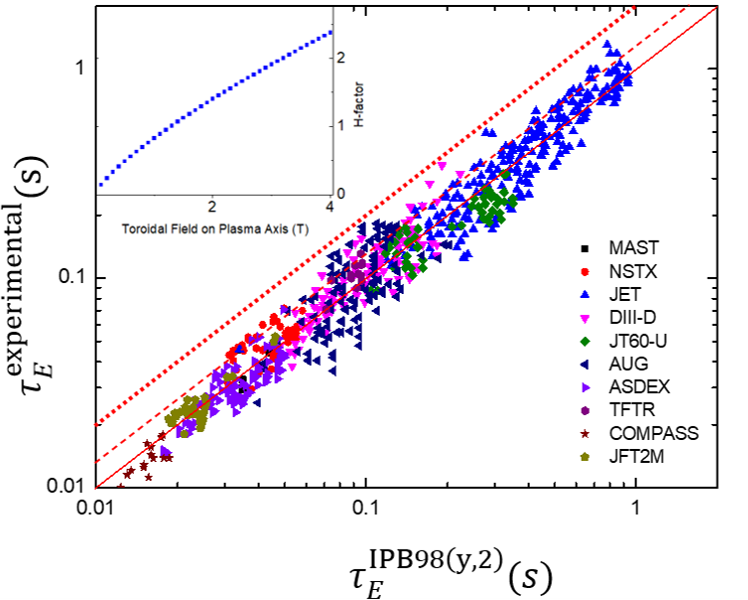}}
    \caption{\label{T_eScalings}A comparison between experimentally measured plasma energy confinement times from various tokamaks and the confinement time predicted by the IPB98(y,2) scaling law and adapted from \cite{mcdonald07}. The solid line is $H_{98} = 1.0$ dashed line is $H_{98} = 1.2$ and the dotted line is $H_{98} = 2.0$. Inset: H-factor as a function of toroidal field on the plasma axis for a tokamak with $I_P$ = 8.20 MA, $P_L$ = 46  \si{\mega\watt}, $n$ = 1.8$\times{10}^{20}m^{-3}$, $M$ = 2.5, $R_{major}$ = 2 m, $A$ = 1.8, $\kappa$ = 3 using the NSTX Gyro-Bohm \cite{buxton19} confinement time scaling.}
\end{figure}

We have considered H$_\text{98}$ = 1.2 as the most likely performance but have also considered the much higher value of H$_\text{98}$ = 1.6 to quantify the possible effects on costs in future from new advanced tokamak designs such as spherical tokamaks \cite{sykes18}. H$_\text{98}$-factor refers to the ratio between observed plasma energy confinement time, $\tau_\text{E}$, and the $\tau_\text{E}^\text{IPB98(y,2)}$ predicted by the ITER Physics Basis ELMy H-mode IPB98(y,2) scaling law \cite{mcdonald07} which is derived from a vast range of tokamaks. A subset of these data are shown in figure \ref{T_eScalings}. Taking a subset of the IPB98(y,2) data  for different reactor geometries can also yield quite different scaling laws and H$_\text{98}$-factors. For example, confinement times of spherical tokamaks appear to have much stronger field dependence \cite{kaye06,buxton19,costley21} than the standard $\tau_\text{E}^\text{IPB98(y,2)} \propto B_\text{T}^\text{0.15}$ e.g. $\tau_\text{E}^\text{MAST} \propto B_\text{T}^\text{1.4}$ in MAST \cite{valovic09}. Extrapolating this to stronger magnetic fields can result in H$_\text{98}$-factors upward of H$_\text{98}$ = 2.0. It is however not clear whether this strong field dependence extrapolates to power plant conditions. The field dependence is linked to strong $\tau_\text{E}$ scaling with plasma collisionality, $\nu_*$, which itself depends on absolute $\nu_*$ \cite{valovic11}: at lower $\nu_*$ the confinement time scaling with $\nu_*$ is reduced. Therefore in higher field tokamaks with reduced $\nu_* (which is \propto B_\text{T}^\text{-4}$) it is unlikely that the strong field dependence will remain. Although ITER is nominally designed with H$_\text{98}$ = 1.0, H$_\text{98} > $1.0 has been observed in a number of existing tokamaks, e.g. DIII-D \cite{wade03}. Indeed ITER is expected to reach H$_\text{98}$ = 1.57 in reversed-shear operation and H$_\text{98}$ = 1.2 in hybrid operation \cite{sips05}. 

\subsubsection{Density, $\beta$ and Safety Factor}
We have chosen a maximum Greenwald fraction at the plasma edge of $f_{GW}^{edge}$ = 0.67 and a peaked density profile such that $f_{GW}^{line-avg}$  = 1.1. The minimum plasma safety factor at the 95 \% poloidal flux surface was set to $q_{95}$ = 3.45. The normalised thermal beta, $\beta_\text{N} \approx$ 2.49 for all reactors. 
These safety limit choices broadly follow the ARC and SPARC philosophies which have ($ f_{GW}^{edge}$ = 0.67, $q_\text{95}$ = 7.2 and $\beta_\text{N}$ = 2.59 \cite{whyte16} and $ f_{GW}^{edge}$ = 0.37, $q_\text{95}$ = 3.4 and $\beta_\text{N}$ = 1 \cite{creely20}, respectively) rather than the EU-DEMO philosophy which will operate closer to stability limits (with $f_{GW}^{edge}$ = 0.8, $q_\text{95}$ = 3.25, and $\beta_\text{N}$ = 2.50 \cite{federici18}). We use the familiar expressions for fusion plasma power \cite{zohm19}: $P_\text{plasma} \propto \beta_\text{N}^2 B_\text{axis}^4 R^3  /  q^2 A^4$ ($B_\text{axis}$ is the magnetic field on the axis of the plasma), and safety factor $q \propto R B_\text{axis}/A^2I_\text{P}$ (for a fixed shaping factor); for a reactor design point with fixed $P_\text{plasma}$: $\beta_\text{N} \propto 1/I_\text{P} B_\text{axis} \sqrt{R}$. Thus going to even higher fields reduces $\beta_\text{N}$. $q$ also scales positively with $B_\text{plasma}$, so larger fields would reduce further the probability of kink disruptions. In addition, $I_\text{P}$ can be increased in tandem with $B_\text{plasma}$, increasing achievable plasma density (as the limiting density $n_G = I_\text{P} / \pi a^2$) whilst maintaining high $q$ and further reducing $\beta_N$. Our calculations show that had we used the higher risk EU-DEMO safety limits for our preferred reactor, it doesn't change things very markedly. The capital cost decreases by 7.9 \%, it decreases the major radius by 4.3 \%, decreases the plasma current by 12.0 \% and increases the field on plasma by 6.1 \%. 

\subsection{Superconductor Operating Temperature} 
We have chosen 4.5 \si{\kelvin} as the operating temperature for all superconducting magnets. All of the reactors in this work use Nb-Ti TF and CS coils at some point during their lifetime, so the cryosystem must be able to cool the magnets to this temperature. Of course, the Nb-Ti training coils (mentioned below) need liquid-helium temperature operation. Even if a REBCO reactor could eventually be operated at 20 \si{\kelvin}, \texttt{PROCESS} shows our preferred choice reactor operating at 4.5 \si{\kelvin} actually has a net capital cost $\approx$ 150 M\$ lower than at 20 \si{\kelvin} (as explained below). For the preferred reactor, the cryogenic cost is 89 kW (compared to the 75 kW cryogenic requirement of ITER \cite{monneret17}). 

Table \ref{Cost_Breakdown} shows the capital cost of our preferred REBCO plant operating at 4.5 \si{\kelvin} including the combined TF (130 M\$) and CS coils' (20 M\$) cable cost also at 150 M\$. Also shown are data for plants where Nb$_3$Sn and Nb-Ti has been used as the superconductor in the highest field regions of the plant. Equivalent reactor power balances are shown in table \ref{Power_Breakdown}. Given the plant with Nb$_3$Sn TF and CS coils is more expensive than the REBCO-based plant (and still made from brittle material) we have not considered it further in this paper. If operation were at 20 \si{\kelvin} REBCO’s critical current density is $\approx$ 1.7 $\times$ lower, which demands larger coils and a larger overall reactor volume, increasing direct costs by 84 M\$. On the other hand, modern cryoplant efficiency scales with temperature roughly as the ideal Carnot cycle (with a base temperature of about 2 K) \cite{stobridge74}.

The direct capital cost of cryoplant scales approximately linearly with cooling power and would be reduced from 88 \$M (as shown in table \ref{Cost_Breakdown}) to 20 \$M. 
Operation at 20 \si{\kelvin} does have the advantage of better REBCO quench mitigation due to the $\approx 3.0~\times$ greater thermal conductivity and $\approx 60~\times$ greater specific heat of RRR = 100 copper at 20 \si{\kelvin} than at 4.5 \si{\kelvin} \cite{simon92}. Though, we expect that with rather modest advances in quench detection and mitigation technologies (e.g. LTS for HTS quench detection \cite{kang21}, acoustic MEMS \cite{takayasu20}, stray capacitance change monitoring \cite{ravaioli20}) operation at 4.5 \si{\kelvin} using REBCO will be straightforward in future. 

\subsection{Tritium Breeding}
The blanket design in all reactors in this work based on the helium-cooled pebble bed (HCPB) \cite{hernandez17} which has greatest breeding potential of the blanket designs under investigation for EU-DEMO \cite{federici19}. We have required the minimum tritium breeding ratio (TBR) to be 1.1 in all reactors. The TBR was set using the in-built \texttt{PROCESS} breeder ratios for given breeder blanket thicknesses as calculated by the \texttt{FATI} (Fusion Activation and Transport Interface) code \cite{shimwell16} for EU-DEMO (to which our designs are similar, to first order, so the model is applicable here). Tritium self sufficiency is required as current tritium supplies could not maintain multiple pilot plant reactors \cite{kovari17}. The TBR cannot be too large, as to avoid an excessive tritium inventory and issues of tritium permeation throughout the reactor. TBR = 1.1 is the widely accepted ratio for a power plant: ``enough but not too much''. 

If the global tritium inventory were markedly increased, a cheaper pilot plant could be built with a lower TBR = 0.9 \cite{NIST21}. Our detailed \texttt{MCNP} (Monte Carlo N-Particle) calculations shows that the 0.53 \si{\metre} thick blanket and 0.25 \si{\metre} thick radiation shield in our preferred reactor each reduce the neutron flux by roughly two or three orders of magnitude (considered below in section 4 and figure \ref{Fluxes}). A TBR of only 0.9 can therefore be generated with a smaller inboard blanket of $\approx$ 0.20 \si{\metre} and outboard blanket of $\approx$ 0.35 \si{\metre} only. However, to maintain the same nuclear heating in the magnets, the radiation shield would need to be thicker by $\approx$ 0.17 \si{\metre} leading to a net reduction in capital cost of $\approx$ 24 \%. The first wall blanket and shield in this case are 70 cm thick in total. This is a very significant reduction (discussed in Section 7.4), but we decided in the end not to pursue it, since although such an approach would still demonstrate (some) tritium breeding, it would be at the cost of losing tritium self-sufficiency (which may unacceptable to investors).

Other breeding blankets are being developed: A water-cooled lithium lead blanket (WCLL) design \cite{martelli17,tassone18} is under consideration for EU-DEMO. The WCLL provides greater radiation shielding than the HCPB (due to neutron capture by the water coolant) whereas the latter has greater breeding potential (due to the inclusion of Be neutron multiplier modules). Other helium-cooled and dual-cooled lithium lead concepts are also under consideration \cite{federici19}. A FLiBe molten salt blanket is being developed in the USA which includes a coolant outlet temperature of up to 930 $^\circ$C \cite{sorbom15}, higher than either the HCPB (650 $^\circ$C) or WCLL (330 $^\circ$C) and may therefore eventually lead to more efficient electricity production. This higher temperature would however require the use of novel, low activation structural materials, EUROfer is limited to 550 $^\circ$C \cite{baluc11}.

\subsection{Reactor Architecture}
\subsubsection{Divertor Constraints and Configuration}
The divertor architecture in all the simulated reactors here is based on the single-null ITER design \cite{pitts19} \cite{morris20}, which is currently the baseline option considered for EU-DEMO \cite{federici18}. The steady-state heat flux onto the divertor was required to be $<$ 6 \si{\mega\watt}/m$^2$, below the maximum steady-state heat flux of $\approx$ 10 \si{\mega\watt}/m$^2$ expected in ITER \cite{pitts19}. This is a conservative constraint, manageable with techniques such as divertor impurity seeding (e.g. in EAST which maintains high H$_\text{98}$ \cite{xu20}) or moving the divertor strike points (e.g. in SPARC \cite{creely20}).In our \texttt{PROCESS} simulations we have allowed the argon impurity fraction to vary, to facilitate reduced power to the divertor through argon ionisation and bremsstrahlung. Other advanced techniques developed for much smaller machines with much higher fluxes are also potentially available including long legged \cite{wigram19,umansky17} or snowflake divertors \cite{ryutov15} but they require additional plasma shaping coils which are exposed to large neutron fluxes, or raise demands (and costs) on the existing coil system \cite{ambrosino19,reimerdes20} (e.g. in ITER, the current through the upper-most and lower-most solenoid modules would have to be increased by more than factor 10 \cite{lackner13} in order to produce a snowflake). In the large, capital-cost minimised machines considered the primary limiting factor preventing smaller sized reactors was the yield stress of the magnet support structural material rather than the heat flux to the divertor; theses divertor configurations were therefore not needed. A hard limit of $P_\text{separatrix}/R_\text{major}$ = 20 \si{\mega\watt}/m$^\text{-1}$ was set, similar to the values of $P_\text{separatrix}/R_\text{major}$ = 17 and 30 \si{\mega\watt}/m$^\text{-1}$ expected for EU-DEMO and J-DEMO respectively \cite{asakura18}. We found that increasing the $P_\text{separatrix}/R_\text{major}$ limit had negligible effect on our cost-optimal designs because they are predominantly magnet stress-limited. 

\subsubsection{Number of Toroidal Field Coils}
All reactors in this work have 18 toroidal field coils and a maximum field ripple at the plasma outboard mid-plane of 6 \% (in following with EU-DEMO designs). Ripple cannot be avoided, but must be kept low in order to reduce ripple-induced drift of trapped particles and associated energy losses \cite{davidson76}. \texttt{PROCESS} runs were performed to ascertain the cost-optimal number of coils for each baseline reactor run in this work. In all cases 18 was the optimum number. The difference in total capital cost between a given reactor with 18 or 20 TF coils was typically quite small: for the preferred reactor the difference was only 0.3 \%. Having a greater number of coils reduces the peak field that each coil must produce (due to the coils be closer together, and the field between them `dipping' less), thereby slightly reducing the coil size and overall reactor volume. Each added coil however increases the cost of the magnet system. 

\subsubsection{Coil Structural Support}
The maximum allowable shear stress (used for the Tresca yield criterion in \texttt{PROCESS}) was set to 660 \si{\mega\pascal} for both the CS and TF coils. This is 2/3 of the yield stress of standard fusion relevant, high strength structural steels \cite{hamada07}. A bucked and wedged (B\&W) coil support structure \cite{titus03,titus16} has been incorporated in all reactors studied here. Performing dedicated \texttt{PROCESS} runs, we found that a B\&W support structure reduces our preferred reactor’s CS coil bore by 13.7 \% (27.9 \si{\centi\meter}), TF coil thickness by 13.4 \% (10.9 \si{\centi\meter}) major radius by 5.6 \% (40.7 \si{\centi\meter}) and capital cost by 400 M\$ compared to a conventional wedged support structure that mechanically isolates the TF coils from the CS coil (as in ITER \cite{aymar02}). In the B\&W support structure, stresses are shared throughout the whole support structure, rather than constrained to the supports of individual coils, reducing the size of the steel support structure required. The TF coils are wedged in a circular vault which bucks onto a low-friction bucking cylinder which itself is in contact with the central solenoid. Such an architecture does however require the use of a bespoke low-friction interfacial material \cite{titus03} and comes at the cost of reduced plasma shaping flexibility, and additional cyclic loading on the TF coils \cite{schultz06} which reduce the fatigue-limited lifetime of the TF coil casing and  has not been accounted for in our calculations. 

The \texttt{PROCESS} stress model is 1-D and only calculates the stress at the inboard mid-plane, from the inner edge of the CS coil to the outer edge of the TF coils. It does not take into account stress peaking along the circumference of a TF coil (due to say toppling forces from interaction with the fields from the PF coils) or within the winding pack. Results are generally consistent with finite element analysis \cite{morris15}.

\subsubsection{Central Solenoid Use and Burn Time}

We have chosen to include both a central solenoid coil and auxiliary heating system for current drive, start-up and plasma heating. To minimise the size of the central solenoid coil, a large 50 \si{\mega\watt} ECRH auxiliary heating current drive was used. This ECRH power follows EU-DEMO \cite{federici18}, which would make it the largest ever built. It would limit any further reduction in the blanket volume (as auxiliary heating systems take up valuable first wall surface area) and hence the tritium breeding ratio and electricity generated. The 50 \si{\mega\watt} ECR system is expected to produce 10 – 15 \% of the plasma current (which has been included in the calculations). It was taken to have a power conversion efficiency $\mu_\text{CD,conv}$ = 0.4, and normalised current drive efficiency of $\gamma_\text{CD}$ = 0.3 - taken from the \texttt{PROCESS} EU-DEMO 2018 baseline values and slightly more conservative than assumed for EU-DEMO \cite{franke14}. \texttt{PROCESS} was then given freedom to vary the inductive and non-inductive current fractions and yielded an inductive (CS and PF coil driven) current fraction of $\approx$ 50 \% (the exact fractions depend on the reactor in question) and a bootstrap current fraction of $\approx$ 40 \%. The CS and PF systems produced $\approx$ half of the total magnetic flux each at all times.

A number of novel plasma start-up techniques have been developed that could in principle reduce the demand on the CS coil, and therefore reduce its size and cost. Helicity injection is a promising family of technologies and have seen implementation in a number of smaller tokamaks \cite{raman14b}. The most powerful system under construction is NSTX-U \cite{raman14a} which is predicted to produce $>$ 400 \si{\kilo\ampere}. Merging compression (MC) has seen some success in spherical tokamaks \cite{sykes01,inomoto15,chung13} and is expected to be used in Tokamak Energy’s ST-40 reactor \cite{gryaznevich17} and produce a 2 \si{\mega\ampere} current. To date MC magnets have been  inside the vacuum vessel which brings with it huge neutron fluxes and the requirement for frequent replacement, reducing reactor availability. Designs that improve the location of the MC magnets will be developed, but we consider this approach too high risk just now. Up to 200 \si{\kilo\ampere} current has also been achieved inductively using the PF coil systems in JT60-U (with supplementation from the lower hybrid current drive system) with 1.9 \si{\weber} flux \cite{ushigome06}, but higher currents must be demonstrated before this technique becomes a practical solution for reactors of the scale considered in this work at this time.

A radio frequency (RF) current drive was chosen for the auxiliary current drive system as it is cheaper, requires less radiation shielding, and consumes a smaller blanket volume than the alternative neutral beam injection system \cite{darbos09} \cite{hemsworth17}. In principle the ECR system could be exchanged for a different 50 \si{\mega\watt} RF current drive option without changing the overall reactor design should ion cyclotron or lower hybrid current drive systems prove more efficient or reliable in future.  For an EU-DEMO-like reactor, at present ECR has the most flexible power deposition which gives the highest current drive efficiency \cite{franke14}. 

\begin{figure}
    \centerline{
    \includegraphics{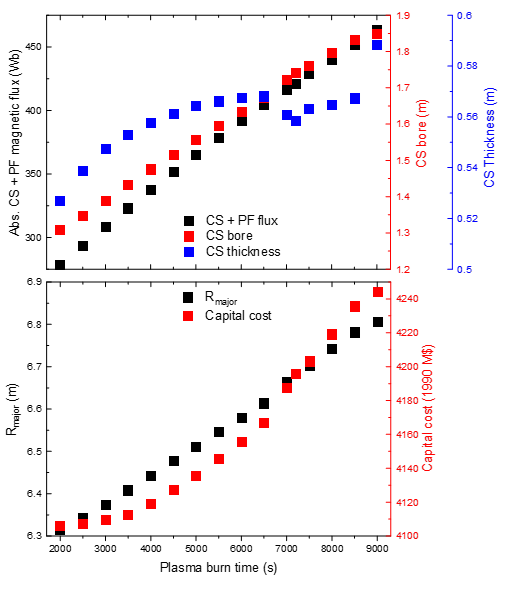}}
    \caption{\label{BurnTime}Top: Total central solenoid coil (CS) and poloidal field coil (PF) flux, central solenoid bore and thickness, and (Bottom:) plasma major radius and plant total capital cost as a function of plasma burn time, of 100 \si{\mega\watt}$_\text{e}$ tokamak power plants with REBCO CS and TF coils and Nb-Ti PF coils (i.e. magnet materials as per the preferred tokamak design) optimised for minimum capital cost. Note that the preferred reactor has a 7200 s plasma burn time.}

\end{figure}

For both the 100 \si{\mega\watt}$_\text{e}$ and 500 \si{\mega\watt}$_\text{e}$ reactors considered here, the capital cost is not very sensitive to burn-time so we have chosen to adopt the EU-DEMO standard of 2 hours \cite{federici18}. The variation in the cost-optimal central solenoid bore, thickness and flux generation as a figure of required plasma burn time and resulting reactor capital cost are shown in figure \ref{BurnTime}.

\section{An Optimised Radiation Shield Thickness}\label{MCNP shield}

In this section we optimise the thickness of the radiation shield. A thinner shield is cheaper and enables more compact reactor designs. However, the shield must be thick enough for both the lifetime of the tokamak to be sufficiently long, and the cryogenic load to be sufficiently small. We start by using state-of-the-art \texttt{MCNP} \cite{goorley12} calculations for the neutron flux spectrum at the first wall for a cost-optimised, H$_{98}$ = 1.2, 100 \si{\mega\watt} REBCO CS and TF and Nb-Ti PF tokamak. We assume that the neutron flux predominantly determines lifetime limit for superconducting materials and hence can be used as a proxy for both the neutron and gamma flux. Then we use \texttt{MCNP} attenuation coefficients derived for neutron flux attenuation through slab geometries, to provide empirical attenuation coefficients for what we call in this paper benchmarking calculations. We have used them here to calculate the lifetime and cryogenic load for a range of simplified tokamak designs using different radiation shield thicknesses. These quick calculations provide a broad view of how changes in the component parts and size of the shield affects the tokamak's performance. Then we progressed to larger \texttt{MCNP} \cite{goorley12} calculations that included the full complexity of the tokamak geometry and both the photon and neutron flux, and optimised the radiation shield thickness more accurately. This finalised the shield thickness of our preferred tokamak. We go on to use the properties for the optimised shield in our cryogenics analysis of the helium coolant mass flow rate required, and in all our subsequent power plant simulations.

\subsection{Neutronics - Thermal Load and Lifetime}

\subsubsection{Benchmarking Calculations}

\begin{figure}
    \centerline{
    \includegraphics[height=12cm,keepaspectratio]{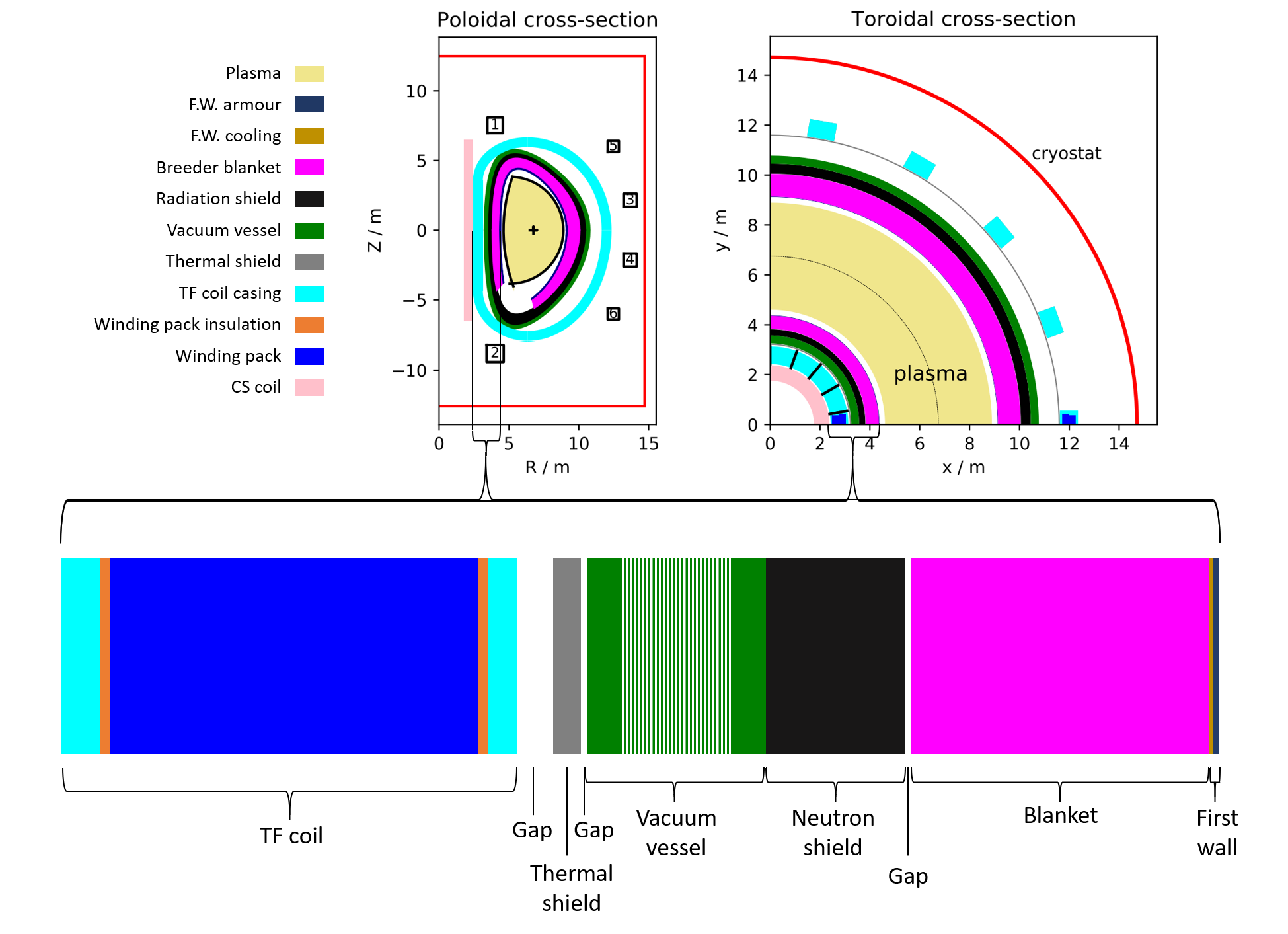}}
    \caption{\label{NewShield}Cross-sections and inboard mid-plane radial build of the preferred reactor (a cost-optimised H$_\text{98}$ = 1.2, 100 \si{\mega\watt} net electricity tokamak with REBCO toroidal field and central solenoid coils) with an optimised 25.0 \si{\centi\meter} radiation shield. Details of the layer constituents can be found in table \ref{ShieldThicknesses}. The inboard blanket is 53 \si{\centi\meter} in radial thickness and based on the EU-DEMO helium-cooled pebble bed design \cite{hernandez17}, guaranteeing a tritium breeding ratio $>$ 1.1.}
\end{figure}

The incident neutron flux density spectrum at the first wall (FW) for our preferred cost-optimised REBCO tokamak $I_\text{FW,RT}(E)$ (n \si{\per\square\centi\meter\per\second}) was calculated using \texttt{MCNP} in terms of $i$ different energy bins of width $\overline{dE_i}$ and average energy $\overline{E_i}$ (and the 175 Vitamin-J energy bin width size distribution \cite{vontobel87} - a choice which does not significantly affect any results in this paper). For all other tokamaks under consideration, the flux density in any $i$th energy bin was then simply given by   

\begin{equation}\label{normalise flux}
    I_\text{FW}(\overline{E_i}) = \frac{P_{Total}}{P_{RT}} I_\text{FW,RT}(\overline{E_i})~,
\end{equation}

where the flux density has simply been scaled by ratio of the total fusion power of the tokamak under consideration to the total fusion power of the preferred cost-optimised REBCO tokamak ${P_{Total}}/{P_{RT}}$. 
The empirical attenuation coefficients used were those calculated using \texttt{MCNP} for neutron transmission through 30 \si{\centi\meter} blocks of mono-material \cite{colling16} and averaged for all fast neutron flux ($E > 0.1 \si{\mega\electronvolt}$). This approach ignores the complexity of the multiple nuclear interactions (discussed below) and simply associates the reduction in energy and flux with a single attenuation coefficient. Table \ref{mus} lists the empirical values for the attenuation coefficients as well as those derived using standard total nuclear cross sections for comparison.  
We then take the thermal load onto the TF coils after passing through all the walls (the first wall, blanket, radiation shield, vacuum vessel and thermal shield) to be   

\begin{equation}\label{eq2}
P_\text{TF}=A\sum_{All~Bins}gI_\text{FW}(\overline{E_i})\times \overline{E_i}\{\prod^\text{All Walls}_{i}\exp[-a\mu_\text{i}x_i](1- \prod^\text{TF Coils}_{i}\exp[-a\mu_\text{i}x_i])\} ~. 
\end{equation}
 where $A$ is the surface area of the first wall. We have introduced two geometrical factors:
 $g$ which accounts for only a fraction of the flux reaching the cryogenic system where $g=\left({R_{major}-R_{minor}}\right)/{R_{major}}$ (i.e. the cryogenic system unlike say the shielding, does not cover the entire surface of the toroid), and $a$ which accounts for the volume of a curved surface being smaller (and therefore attenuating less on the inner leg of the important TF coils) than a slab where $a=1-{t_{All~Walls}}/{2.r_{All~Walls}}$. For the preferred reactor, $R_{major}=6.75$ \si{\meter} and $R_{minor}= 2.14$ \si{\meter} (cf Table 1). Also ${t_{All~Walls}}= 1.238$ \si{\meter} taken for the first wall, breeder blanket, radiation shield and vacuum vessel given in Table \ref{ShieldThicknesses} and $r_{All~Walls}= 3.383$ \si{\meter} from Figure \ref{NewShield}, so $g=0.682$ and $a=0.817$. Because these corrections appear in exponential functions, they significantly improve the agreement between the benchmarking calculations and the \texttt{MCNP} calculations provided below.   

 To calculate the lifetime of the tokamak, we note that neutron flux density initially increases $J_\text{c}$ in superconductors, due to an increase in the density of flux pinning sites \cite{fischer18}, but eventually causes a sharp irreversible decrease after a fluence of $\approx 3.9 \times 10^{22}$ fast neutrons m$^{-2}$, for $E_{neutron} > 0.1$ \si{\mega\electronvolt} . We have used this fluence threshold (aka the Weber dose limit \cite{weber11}) to calculate the magnet lifetime of the toroidal field (TF) coils $\tau_\text{TF}$ (s), where
\begin{equation}\label{eq3}
\tau_\text{TF}=\frac{3.9 \times 10^{22}}{\sum_{All~Bins} gI_\text{FW}(\overline{E_i})\{\prod^\text{All Walls}_{i}\exp[-a\mu_\text{i}x_i](1- \prod^\text{TF Coils}_{i}\exp[-a\mu_\text{i}x_i])\}}.
\end{equation}

To validate these benchmarking calculations, we first input the radial build dimensions and fusion plasma power for ITER \cite{aymar02} and compare the values obtained to more detailed neutronics calculations \cite{richard14}. With $P_{ITER,plasma}$ = 500 \si{\mega\watt}, a first wall surface area of 610 \si{\square\meter}, $R_{ITER,major}=6.20$ \si{\meter}, $R_{ITER,minor}= 2.00$ \si{\meter},  ${t_{ITER,All~Walls}}= 0.808$ \si{\meter} and $r_{All~Walls}= 3.817$ \si{\meter} (as shown in Table \ref{ReactorComparison}), our benchmarking calculations yield a TF coil nuclear heating of 32.8 \si{\kilo\watt}, within just a factor of two of the expected range of 14 - 18 \si{\kilo\watt} \cite{richard14} (The calculated magnet lifetime for ITER is 23.6 full-power years). 
Given the agreement, we then changed the radiation shield to be tungsten carbide and used \texttt{PROCESS} to vary the thickness of the components of the radial build and found the first approximate design of the preferred tokamak (the data for this initial design are listed in table \ref{ShieldThicknesses}). Having found the first approximate design for the preferred tokamak, \texttt{MCNP} was then used to finalise the radiation shield thickness.

\subsubsection{\texttt{MCNP} Calculation}

\begin{figure}
    \centerline{
    \includegraphics[height=9cm,keepaspectratio]{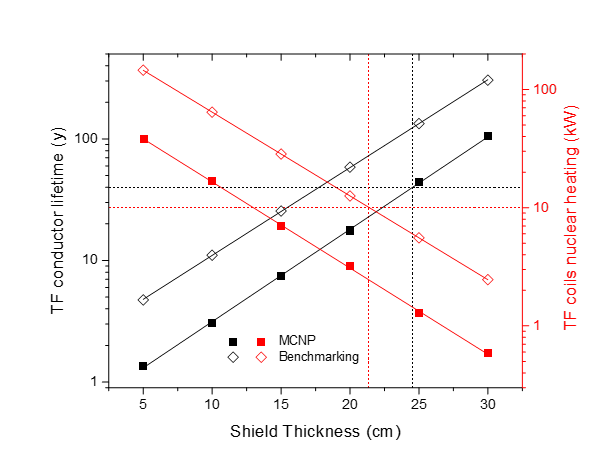}}
    \caption{\label{Xplot} Nuclear heating in the TF coil system and number of full-power operation years until the Weber dose limit \cite{weber11} is achieved for the preferred reactor (a cost-optimised H$_\text{98}$ = 1.2, 100 \si{\mega\watt} net electricity tokamak with REBCO toroidal field and central solenoid coils) as a function of the thickness of its tungsten carbide radiation shield as calculated by \texttt{MCNP} (closed squares) and the benchmarking calculation (open diamonds). The material layers between the plasma and the TF coil are given in table \ref{ShieldThicknesses}. The dotted black lines indicate the minimum 40 year conductor lifetime limit (and corresponding minimum shield thickness as calculated by \texttt{MCNP}). The dotted red lines indicate the maximum 10 \si{\kilo\watt} heating limit on the TF system (and corresponding minimum shield thickness as calculated by our benchmarking calculations).}
\end{figure}

The \texttt{MCNP} code is a dedicated numerical solver that considers the progressive creation and loss of approximately 4000 isotopes, via decay and nuclear reactions. These calculations include the complexity of considering flux in all directions, and the specific geometry of the component structures of the tokamak \cite{goorley16,mashnik11}. It is used here to calculate the changes in the neutron flux and gamma flux as they pass through the component walls and magnets of the tokamak. The calculations do not include changes in composition or microstructure that affect mechanical properties, such as embrittlement or swelling \cite{gilbert12,chakin06}, nor do they include changes in transport properties, such as thermal or electrical conductivity \cite{hofmann15}, or magnetic properties. Having used the benchmarking calculations to identify the first approximate optimal design for a cost-optimised REBCO tokamak, we repeated the nuclear heating and superconductor lifetime calculations using \texttt{MCNP} near the optimal shield design. The space for the tungsten carbide radiation shield was set as a 30 \si{\centi\meter} block and split into six, 5 \si{\centi\meter} thick sections. The sections were successively set as void regions starting from the plasma facing side, and the neutron and photon flux density spectra were calculated for materials throughout the entire tokamak together with the lifetime and cryogenic load on the TF coil system, as shown in figure \ref{Xplot}. Also shown are equivalent benchmarking values. The \texttt{MCNP} calculated lifetimes and TF coil nuclear heating as a function of shield thickness are $\approx$ 3-4 $\times$ and  $\approx$ 4 $\times$ lower than the corresponding benchmarking values for a given radiation shield thickness. Further corrections can be added to the benchmarking calculations by accounting for the higher $> 10$ \si{\mega\electronvolt} neutron flux and lower $0.1 - 10$ \si{\mega\electronvolt} neutron flux at the magnets, than \texttt{MCNP} calculations give, and which lead to the total fast neutron flux being a little lower (resulting in a longer superconductor lifetime) and the total power deposited in the magnets being a little larger (resulting in a larger nuclear heating). 

\begin{figure}
    \centerline{
    \includegraphics{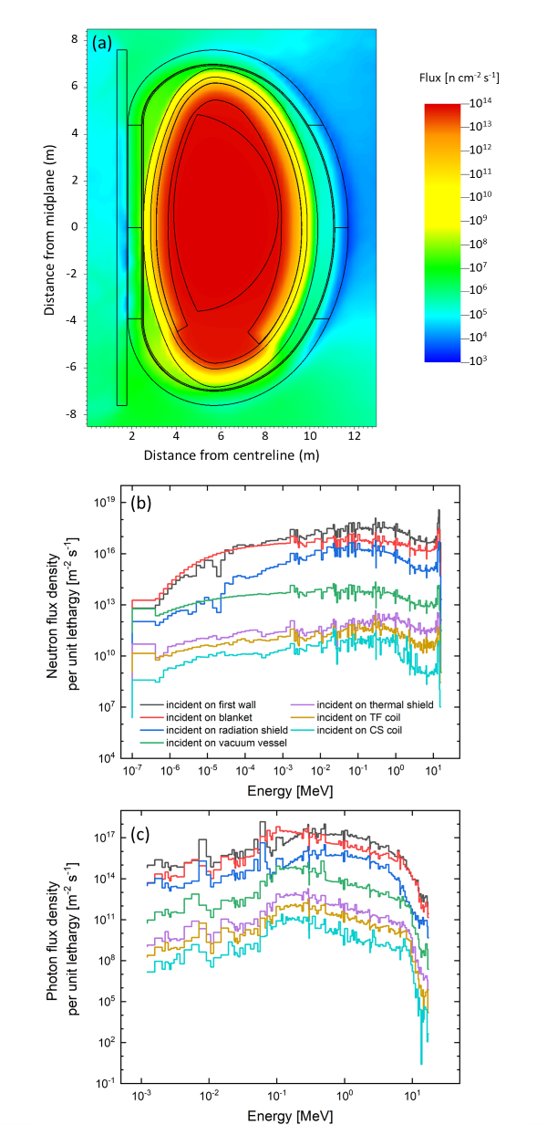}}
    \caption{\label{Fluxes}\texttt{MCNP} calculated (a) fast neutron flux through the poloidal cross-section of the preferred reactor (a cost-optimised H$_\text{98}$ = 1.2, 100 \si{\mega\watt} net electricity tokamak with REBCO toroidal field and central solenoid coils) with an optimised 25.0 \si{\centi\meter} radiation shield detailed in the right hand column of table \ref{ShieldThicknesses}. (b) The neutron flux spectrum and (c) photon flux spectrum as a function of distance into inboard mid plane of the preferred reactor with a 25.0 \si{\centi\meter} radiation shield. These spectra were converted to flux density per unit lethargy by multiplying the spectral histogram fluxes by the ratio between the energy bin average energies and the bin widths.}
\end{figure}

\begin{figure}
    \centerline{
    \includegraphics[height=9cm,keepaspectratio]{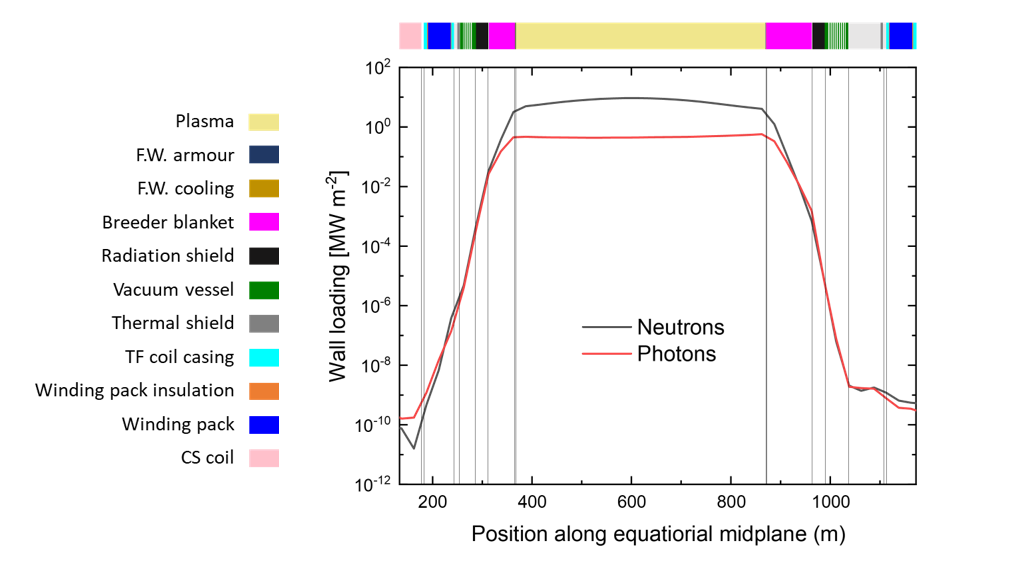}}
    \caption{\label{WallLoading}Neutron and photon induced wall loading along the mid-plane within the preferred reactor (a cost-optimised H$_\text{98}$ = 1.2, 100 \si{\mega\watt} net electricity tokamak with REBCO toroidal field and central solenoid coils) with an optimised 25.0 \si{\centi\meter} radiation shield. The radial positions of the central solenoid coil (light pink), toroidal field coil legs (blue), vacuum vessel (green), radiation shield (black) and blanket (deep pink) are shown.}
\end{figure}

The neutron and photon spectra as a function of depth into the reactor wall at the inboard mid-plane are shown in figure \ref{Fluxes}. The data are presented as flux density per unit lethargy (i.e. flux density in the $i$th bin, divided by the $i$th energy bin width, and multiplied by the average energy in the bin) versus energy. This form is independent of the details of how the bins are discretised and enables direct comparison with for example Weber \cite{weber11} who finds a peak value of $\approx 4 \times 10^{12}$ n \si{\per\square\meter\per\second} at the magnet location, that is similar to the peak flux of $2.3 \times 10^{12}$ n \si{\per\square\meter\per\second} incident on the TF coils shown in figure \ref{Fluxes}. We note that the gamma flux is generally lower than the neutron flux although to our knowledge there are no reports of how this may affect the lifetime of the superconductors (cf Section 5.6). Neutron and photon (power) wall loading are shown in figure \ref{WallLoading}. 

The optimal tungsten carbide radiation shielding thickness was calculated  using \texttt{MCNP} to be 24.5 \si{\centi\meter}, based on a 40 year superconductor lifetime criterion. With this shield the combined nuclear heating on the TF coils was very low, only $\approx$ 1.4 \si{\kilo\watt}. We note that if we had chosen to reduce the lifetime to just 3 years, the shielding would have reduced to 9.8 \si{\centi\meter}, but at the price of the nuclear heating increasing to a large value of 17.2 \si{\kilo\watt} and the cost reducing by less than 5 \%. We did not pursue this option further. A 25.0 \si{\centi\meter} shield was employed for all of our further \texttt{PROCESS} calculations. A breakdown of the resulting optimised reactor radial build is shown in table \ref{ShieldThicknesses} and figure \ref{NewShield}. The approach we have adopted here has identified the important properties of our preferred choice of reactor using state-of-the-art \texttt{MCNP} calculations. The benchmarking data in Table \ref{mus} demonstrates that the neutron and gamma flux typically reduces by an order of magnitude every 15 \si{\centi\meter}. The detailed \texttt{MCNP} calculations confirm that unless one is going to embark on regular component replacement, the range of radiation shield thicknesses available to the fusion engineer is limited, given the TBR requirements, and that currently, the optimum is determined by the lifetime of the best available brittle superconductors. 

\subsection{Cryogenic flow - Benchmarking}

In this paper, we assume that the cryogenic heat load is broadly constant throughout a plasma pulse and distributed evenly throughout each cooling channel. The benchmarking thermal calculations only consider the TF coils (whereas the detailed \texttt{PROCESS} calculations consider the entire magnet system - TF, CS and PF coils). The temperature of the superconductor can be estimated using Newton's law of cooling \cite{Newton1701}
\begin{equation}\label{eq4}
    T_\text{sc}(x) = T_\text{coolant}(x) + \Delta T_\text{coolant-sc} = \left(T_\text{coolant}^\text{inlet} + \frac{Qx}{\dot{m}L_\text{channel}c_p}\right) + \frac{Q}{W_pL_\text{channel}h} ~,
\end{equation}
where $\Delta T_\text{coolant-sc}$ is the difference between the temperature of the coolant and that of the superconductor, $Q$ is the heating load (in Watts) in each cooling channel, $L_\text{channel}$ is the cooling channel length, $W_p$ is the cooling channel wetted perimeter, $\dot{m}$ is the coolant mass flow rate, $x$ is the distance along the cooling channel,
%$p$ is the coolant pressure,
$c_p(T)$ is the coolant specific heat capacity per unit mass, and the heat transfer coefficient, $h(T,p)$, can be derived from the Dittus-Boelter equation, written in terms of the Nusselt number, $Nu$, \cite{varin20}; 
\begin{equation}\label{eq5}
    Nu = \frac{hD_h}{\kappa} = 0.023 Re^{0.8}Pr^{0.4}~,
\end{equation}
where $D_h$ is the cooling channel hydraulic diameter, $\kappa(T,p)$ is the coolant thermal conductivity $Re(T,p) = \dot{m} D_h/ \mu(T,p) A_{coolant}$ is the coolant Reynolds number, $A_{coolant}$ is the coolant cross section, $Pr(T,p) = \mu(T,p) c_p(T,p)/\kappa(T,p)$ is the coolant Prandtl number and $\mu(T,p)$ is the coolant dynamic viscosity. From the maximum pressure drop allowed, the maximum mass flow rate can be calculated using the Darcy-Weisbach equation \cite{varin20}
\begin{equation}\label{eq6}
\Delta P(x) = \frac{f_d x \rho_V {\langle v \rangle}^2}{2D_h}
\end{equation}
where $\rho_V$ is the density, $\langle v \rangle = \dot{m}/\rho A_{coolant}$ is the mean coolant flow velocity and the Darcy friction factor $f_d$ can be expressed in terms of the Reynolds number and void fraction in the cable -  $V_{coolant}$ as \cite{varin20}
\begin{equation}\label{eq7}
f_d = \frac{19.5 / Re^{0.7953} + 0.0231}{V_{coolant}^{0.742}}~.
\end{equation}
We validate this benchmarking approach by considering the JT60-SA tokamak and the materials properties used in Table \ref{Cryoproperties}. Using $A_{coolant} = 1.27\times10^{-4} \si{\square\meter}$, $D_h = 4.57\times10^{-4}$ \si{\meter}, $\dot{m} = 3.5$ \si{\gram\per\second}, $V_{coolant} = 0.32$, $T_\text{coolant}^\text{inlet} =$ 4.4 \si{\kelvin}, an inlet pressure of 5 \si{\bar}, $L_\text{channel}$ = 123.3 \si{\meter} (5 double pancakes per TF coil, each of length 296 \si{\meter} \cite{JT60mags} and 12 cooling channels per TF coil \cite{varin20}), the time averaged heat load on each coolant channel is 12.1 \si{\watt}. Equations \ref{eq4} - \ref{eq7} yield a pressure drop of 0.9 \si{\bar} and helium outlet temperature for each cooling channel of 5.2 K, which compares favourably to more detailed calculations of 1.1 \si{\bar} and $\approx$ 4.8 \si{\kelvin} \cite{varin20}.

For our preferred choice reactor, PROCESS gave 18 TF coils with a total cable cross section of 39.6 \si{\square\centi\meter} and an inner (square) cross section of 23.6 \si{\square\centi\meter}. We have set the number of cooling channels per TF coil to be 10 (note JT-60SA has 12 and ITER has 14), which given there are 100 turns per TF coil each of 36.4 \si{\meter}, leads to $L_\text{channel}$ = 364 \si{\meter}. A 20 \% conductor void fraction \cite{Slade19} then sets the cooling channel hydraulic diameter in the superconducting cable to be 2.45 \si{\centi\meter} (with $V_{coolant} = 1.0$ within this channel). Setting the inlet temperature to 4.5 \si{\kelvin}, inlet pressure to 5 \si{\bar} and limiting the pressure drop along the coolant pipe to no more than $\Delta P$ = 1.0 \si{\bar} sets an upper limit on the fluid mass flow rate of $\approx$ 132 g s$^{-1}$ (much larger than in JT-60's 3.5 g s$^{-1}$ because the channel is much wider and is unobstructed by conductor strands, with commensurately less drag). Including the TF coil winding pack circulator work, AC losses and static heat loads the total \texttt{PROCESS} calculated heat load is 89 \si{\kilo\watt} and the TF coil coolant outlet temperature is 4.6 \si{\kelvin}. Hence we conclude that the cryoplant performance required by the preferred choice reactor is met using existing tokamak cryoplant systems.

It is interesting to consider whether, if operation were required at 30 \si{\kelvin}, a different cryogen, would be preferred. Here we rule out hydrogen and oxygen mixes to avoid unnecessary additional safety considerations and just consider neon. Under 5 \si{\bar} pressure, supercritical helium at 30 \si{\kelvin} has $\approx$ 6 \% of its density and  $\approx$ 120 \% of its dynamic viscosity at 4.5 \si{\kelvin} \cite{TPFS}. If we maintain the 1.0 \si{\bar} pressure drop, the mass flow rate reduces to 30 g s$^{-1}$ resulting in an outlet temperature of 30.3 \si{\kelvin}. At 30 \si{\kelvin} and 5 \si{\bar}, liquid neon has a density $\approx$ 9 $\times$ greater and a dynamic viscosity $\approx$ 25 $\times$ greater than He at 4.5 \si{\kelvin}. A 1.0 \si{\bar} pressure drop leads in this case to a mass flow rate of 380 g s$^{-1}$ and an outlet temperature of 30.1 \si{\kelvin}. Hence, at 30 \si{\kelvin} liquid neon only slightly outperforms supercritical helium as cryocoolant and so is not required/considered further.  

\section{High Field Superconductors}\label{Superconductors}

Here we consider the important high-field superconductors that can enable commercial magnetically confined fusion: 

\subsection{Fusion Relevant Superconductors}

\subsubsection{Nb-Ti~}

The Nb-47wt.\%Ti alloy \cite{liu10} is the most important commercial superconducting material. It has been optimised for maximum critical current density between $\approx$ 4 \si{\tesla} and 6 \si{\tesla} for MRI and accelerator magnet applications \cite{lee12}. Its relatively low upper critical field ($B_\text{c2}$(4.2 \si{\kelvin}) $\approx$ 10 \si{\tesla} \cite{chislett-mcdonald20a}) means that it has only been used for the poloidal field coils in next generation fusion reactors such as ITER \cite{sborchia11} and EU-DEMO \cite{federici18} (though it is being used for the TF coils in JT60-SA \cite{shirai17}).

\subsubsection{Nb$_3$Sn}

Nb$_3$Sn  is a brittle intermetallic compound with $B_\text{c2}$(4.2 \si{\kelvin}) $\approx$ 20 $\si{\tesla}$ \cite{taylor05} which has long made it the material of choice for applications when $>$ 10 \si{\tesla} fields are required. The Nb$_3$Sn superconducting matrix can also include tantalum and titanium \cite{cheggour10} (to increase the upper critical field) or hafnium \cite{balachandran19} dopants (for improved $J_\text{c}$ at fields above 15 \si{\tesla}). Nb$_3$Sn cables are broadly produced in one of two ways: Wind \& React where unreacted cables are jacketed and wound into a coil which then undergoes heat treatment; or React \& Wind where the cables are heat treated and then wound into a coil \cite{bruzzone18b,sedlak17}. When using the former process one has to be careful about the fracture of the Nb$_3$Sn filaments and consequent degradation of the cables’ critical current during manufacture \cite{bruzzone11,uglietti18} due to the different thermal expansions of the cable jacket and superconducting filaments. The latter method avoids this issue, but the reacted cable can only be used to produce magnets with large bending radii. Nb$_3$Sn is not considered in detail in this work because we have found almost always that a cost minimised Nb$_3$Sn reactor has a larger capital cost than an equivalent REBCO reactor, as shown in table \ref{Cost_Breakdown}. However as discussed below in section \ref{Gradedcoils}, Nb$_3$Sn could still have a role to play in cost optimising graded coils where different superconductors are used within the same winding pack.  

\subsubsection{REBCO}
The exciting new results from MIT which achieved a field of $>$ 20 \si{\tesla} \cite{CFSMagnet,CFS_video} at elevated temperatures (20 \si{\kelvin}) in a fusion relevant coil, demonstrate REBCO cables are on a fast track to fusion applications \cite{hartwig20}. REBCO's $\approx$ 90 \si{\kelvin} critical temperature \cite{branch20} allows for large temperature margins in cable design and higher operation temperature that reduces cryo-power requirements. However there is more work to be done to demonstrate reliability - it is a ceramic oxide material, that is prone to brittle fracture under tensile strain $>$ 0.3 - 0.7 \% \cite{osamura16,barth15} and tape delamination under cyclic loading \cite{bruzzone18a,maeda14}. Although stable against quenches, quench protection and detection are more demanding than in low temperature superconductors due to REBCO's low normal zone propagation velocities \cite{lacroix13} and the low thermal conductivity in the tapes ($\approx$ 100 - 600 \si{\watt\per\meter\per\kelvin} at 20 \si{\kelvin} and zero field \cite{bonura15}).

\subsubsection{New fusion-focused high-field superconducting alloys}

Other Nb-Ti based alloys have been produced with larger upper critical fields than commercial Nb-Ti. Indeed the record upper critical field at 4.2 \si{\kelvin} is held by a quaternary alloy Nb 38.5\%wtTi 6.1\%wtZr 24.3\%wtTa with $B_\text{c2}$(4.2 \si{\kelvin}) $\approx$ 13 \si{\tesla} \cite{horiuchi73,collings86}. Although the alloy is not produced commercially, its higher $B_\text{c2}$ and ductility make it a obvious candidate material to optimise for future high-field fusion coils. In this work we have completed cost calculations using both the commercially available Nb-Ti used in ITER, and quaternary Nb-Ti (with the implicit assumption that fusion on an industrial scale would provide the commercial driver for quarternary Nb-Ti if required, at a similar cost to current commercial Nb-Ti). These calculations demonstrate that in fusion magnets, unlike accelerator magnets, it is is the low resistance rather than the high $J_c$ values that is required from superconducting materials. This points to future work (beyond the scope of this paper) developing fusion-focused high $B_{c2}$ superconductors that may be new alloys, or perhaps exploit reduced dimensionality to produce high $B_{c2}$ \cite{Tinkham96} in say artificial multilayer alloys that bring the huge potential advantages of lower cost, more straightforward robotic handling, higher radiation tolerance and higher strength than brittle materials and hence could displace high temperature superconductors.            

\subsection{Critical current density - field, temperature and strain dependence}
\begin{figure}
    \centerline{
    \includegraphics[height=9cm,keepaspectratio]{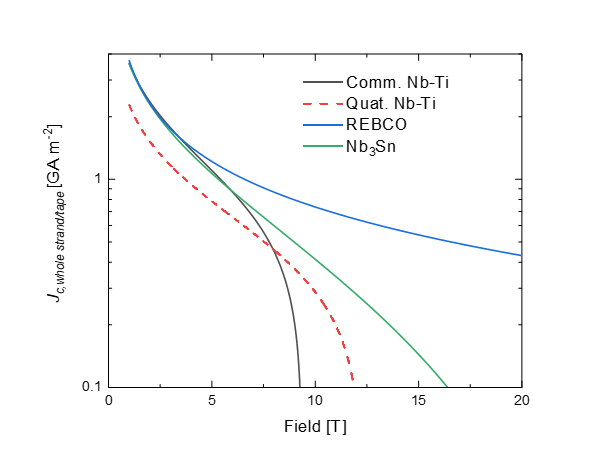}}
    \caption{\label{JcPlots}Whole strand/tape critical current density of commercial ITER specification Nb-Ti (Comm. Nb-Ti), quaternary (Quat.) Nb-Ti,  internal tin Nb$_3$Sn, and REBa$_2$Cu$_3$O$_{7}$ (REBCO where RE: rare-earth) at 4.5 \si{\kelvin}, used in this work. Quaternary Nb-Ti is not commercially available (but could be optimised for fusion applications).}
\end{figure}

Updated \texttt{PROCESS} subroutines for Nb-Ti, Nb$_3$Sn and REBCO have been used throughout this investigation, primarily based on data collected at Durham University. All data in table \ref{JcScaling} correspond to the whole strand or whole tape critical current density (aka the engineering current density). When modelling the low temperature superconductors using \texttt{PROCESS} the cable conductor fraction of copper is 69 \% and of superconductor is 31 \%. The cable conductor helium void fraction is 33 \% (similar to the ITER cables \cite{devred14}). For REBCO, we have assumed the cable is fabricated with stacked tapes (similar to \cite{Slade19}) and has a helium void fraction of 20 \%. The operating current was in all cases set to 100 \si{\kilo\ampere} and limited to 50 \% of the cable critical current in all cases. 
The parameterisations for the critical current density, $J_\text{c}$, of Nb-Ti, quaternary Nb-Ti, Nb$_3$Sn and REBCO are based on a standard scaling law \cite{chislett-mcdonald20a,keys03} itself based on the well-established Ginzburg-Landau theory for the high field properties of superconductors \cite{dew-hughes74,wilson86,tilley90}:
\begin{equation}\label{eq1}
J_\text{c}(B,T,\epsilon) = A^* \left[T_\text{c}^*(1-t^2)\right]^2\left[B_\text{c2}^*\right]^\text{n-3}b^\text{p-1}(1-b)^q~, 
\end{equation}
 where $B$ is the applied magnetic field, $T$ is the temperature, $\epsilon$ is the strain, $A^*$ is a constant, $T_\text{c}^*$ is the critical temperature, $B_\text{c2}^*$ is the upper critical field, $b = B/B_\text{c2}^*$ and $t = T/T_\text{c}^*$. $A^*$, $T_\text{c}^*$ and $B_\text{c2}^*$ have been given their standard literature values \cite{taylor05,lu08} found for an operating temperature of 4.5 \si{\kelvin} and an applied strain of -0.5 \% (equivalent to an intrinsic strain of -1.0\%). These strain values were fixed for all conductors, representative of typical cryogenic pre-strain (a compressive strain of -0.58\% is expected for ITER conductors \cite{nijhuis05}). Data from measurements on ITER specification commercial Nb-Ti have yielded the fit parameters detailed in table \ref{JcScaling} \cite{chislett-mcdonald20a}. As mentioned, although quaternary Nb-Ti has not been commercialised or produced in wire form, we have addressed its potential by using the literature values of $B_\text{c2}$(4.2 \si{\kelvin}) and $T_\text{c}$ for bulk materials reported in \cite{horiuchi73,collings86}. All other fitting parameters for this quaternary material were assumed to be the same as for commercial Nb-Ti. The REBCO scaling parameters were taken from previous measurements on SuperPower tapes \cite{branch20,braccini10}. A comparison between the overall strand and tape critical current densities of commercial Nb-Ti, quaternary Nb-Ti and REBCO at 4.5 \si{\kelvin} is shown in figure \ref{JcPlots}.

\subsection{Costing Superconductors}\label{S/C_costs}
The superconductors have been costed using the industry standard units \cite{cooley16} of \$/\si{\kilo\ampere\meter} where 
\begin{equation} \label{cost_equation}
    \text{Cost}(B,T) = \text{Cost}(B_\text{ref}, T_\text{ref}) \times \frac{J_\text{c}(B_\text{ref}, T_\text{ref})}{J_\text{c}(B,T)}
\end{equation}
where $B_\text{ref}$ and $T_\text{ref}$ are reference conditions at which cost is usually quoted. Increasing the tape/wire current density for a factor of $n$ decreases the cost by the same factor of $n$ (assuming that manufacturing costs remain unchanged). We have used a cost of 1.7 \$/\si{\kilo\ampere\meter} (6 \si{\tesla}, 4.2 \si{\kelvin}) for both commercial Nb-Ti and quaternary Nb-Ti strands, and 8.0 \$/\si{\kilo\ampere\meter} (6 \si{\tesla}, and 4.2 \si{\kelvin}) for Nb$_3$Sn strands \cite{lee15} (in 2021 costs). Currently, REBCO tapes are priced at $\approx$  80 \$/\si{\kilo\ampere\meter} (6 \si{\tesla}, 4.2 K) with the aim to reduce this to 30 \$/\si{\kilo\ampere\meter} (6 \si{\tesla}, 4.2 K) in the near future \cite{cooley16}. Increased demand could reduce this even further to 10 \$/\si{\kilo\ampere\meter} (6 \si{\tesla}, 4.2 \si{\kelvin}) \cite{cooley16,yamada15}. Here REBCO costs of 10 \$/\si{\kilo\ampere\meter} and 30 \$/\si{\kilo\ampere\meter} have been used for the H$_\text{98} = 1.6$  and H$_\text{98} = 1.2$ reactor studies respectively and are representative of the market prices of the superconducting strands/tapes, which are typically 10$\times$ \cite{cooley16} or even 20 - 35$\times$ the raw material costs \cite{bruzzone21}.

The trustworthiness of this cost model was ascertained in three ways: (1) the cost model was applied to \texttt{PROCESS} test-cases (such as the 2018 EU-DEMO baseline model). (2) Runs were performed using the \$/kg model and then \$/\si{\kilo\ampere\meter} model. Typically, magnet costs from the new cost model differed from those of the original model by $<$ 20 \%, and were as expected in more extreme cases (such as for a REBCO cost of 0.025 \$/\si{\kilo\ampere\meter} as in section \ref{Future developments}). (3) The relative costs between systems were compared to those from independent studies of other tokamaks (such as ARC \cite{sorbom15}, ITER and EU-DEMO \cite{mitchell21}). Relative costs between plant components are broadly in line with what would be expected for the size of reactors investigated. 

\subsection{Graded and Sectioned Coils} \label{Gradedcoils}
\begin{figure}
    \centerline{
    \includegraphics[height=8cm,keepaspectratio]{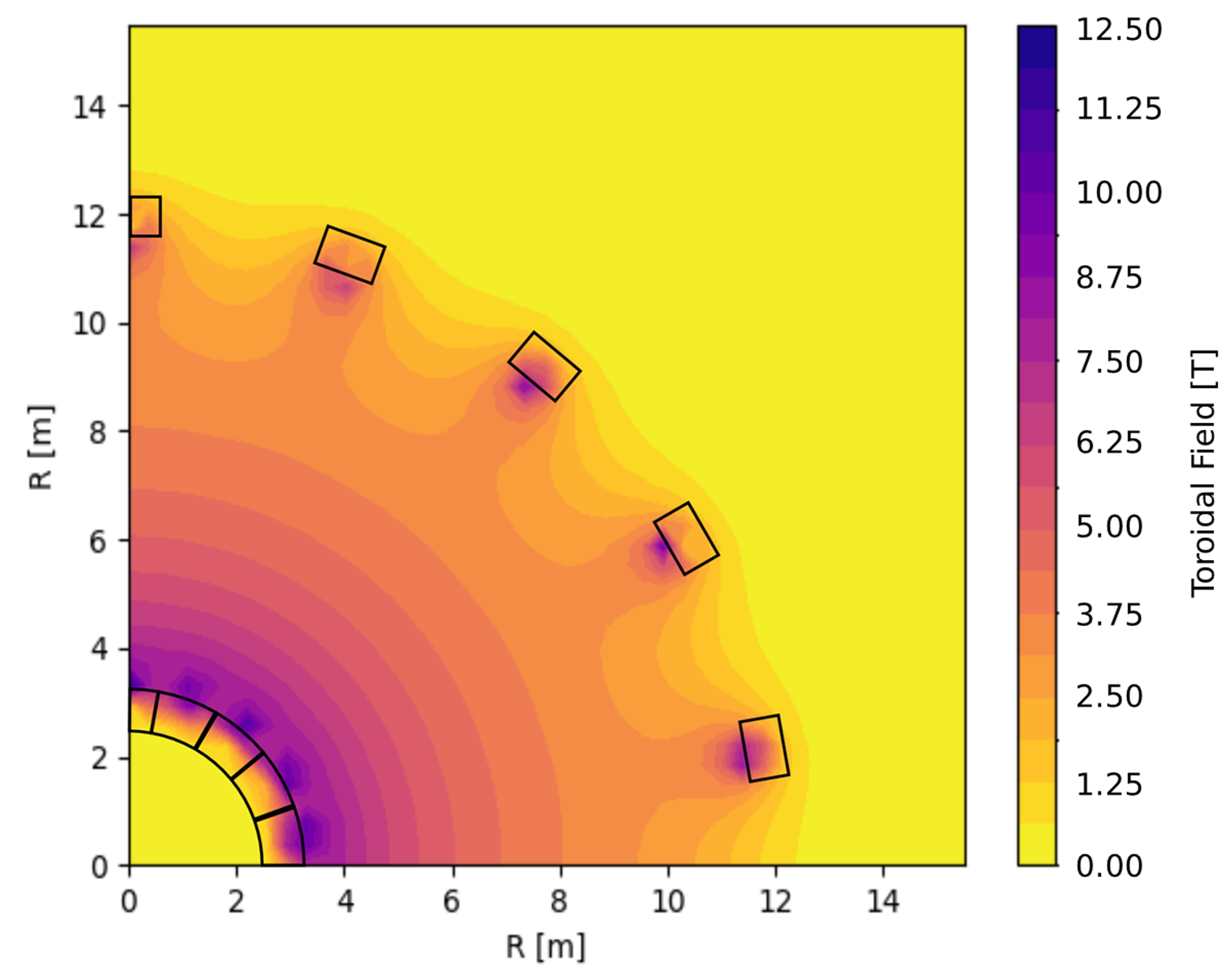}}
    \caption{\label{BLUEPRINT} Toroidal magnetic field through the midplane of the preferred reactor (a cost-optimised H$_\text{98}$ = 1.2, 100 \si{\mega\watt} net electricity tokamak with REBCO toroidal field and central solenoid coils).}
\end{figure}

Here, we have used the most simple winding pack design which retains the same superconductor cross section along the entire cable length (as determined at the peak field on coil) \cite{sborchia08}. However in the regions of the magnets where the field is low, the superconductor operates well below its critical current density and  one can consider graded coils in which the cross section of the cables are reduced \cite{sedlak17,Savoldi17}. Further cost reductions follow when cheaper superconductors are used in the outer parts of the winding pack e.g. Nb-Ti in the low field regions, Nb$_3$Sn in the middle of the winding pack and REBCO in the high field regions \cite{wesche18a}. Figure \ref{BLUEPRINT} shows the clear benefit using graded TF coils. The inboard side of the TF outer leg sees fields $\approx$ 30 \% lower than the maximum on-coil field (at the outboard side of the TF inner leg), and the outboard side of the outboard leg sees fields 75 \% lower than the maximum on-coil field. The toppling forces are also localised, meaning that grading cable conduit thickness is also beneficial. Taking the example of the central solenoid coil in \cite{sarasola20} and using a REBCO cost of 30 \$/\si{\kilo\ampere\meter} (6 \si{\tesla}, 4.2 \si{\kelvin}), we calculate the graded multi-superconductor solenoid to have a materials cost $\approx$ 21 \% cheaper (equivalent to an overall capital cost reduction of just 0.3 \%) than the ungraded REBCO-only solenoid. 

As well as traditionally graded coils, the field-on-coil data of figure \ref{BLUEPRINT} show that given we need remountable magnets to enable timely repair, we can also consider sectioned coils, perhaps with a half-phi design \cite{hampshire18}: coils where the inner and outer coil limbs are based on different superconductors \cite{chislett-mcdonald19, hampshire18}. For the preferred reactor TF coils, the field on the outer limb is below 8 \si{\tesla}, making Nb-Ti the obvious choice. Nb-Ti at 8 \si{\tesla} has a cost in \$/\si{\kilo\ampere\meter} $\approx$ 16 $\times$ lower than that of REBCO at 12.5 \si{\tesla}, so adopting Nb-Ti outer limbs would reduce the preferred reactor’s TF coil direct cost by $\approx$ 16 \% (reducing the  reactor's overall capital cost by $\approx$ 2 \%). 

\subsection{Stress-limited, $J_\text{c}$-limited and $B_\text{c2}$-limited magnets} \label{StressandJcMagnets}

In general, superconducting magnets can be: stress-limited, in which case the Lorentz force induced stresses are close to the yield-stress of the component magnet material; $J_\text{c}$-limited, where the current density in the superconductor is sufficiently high to overcome flux pinning and the material may become resistive. Here we also consider a type of $J_\text{c}$-limited, that we call $B_\text{c2}$-limited. In this case, the operating field is close to the upper critical field of the superconductor so the description makes clear that increasing $B_\text{c2}$ will significantly affect the operating field achievable - equivalent to the critical current density being low because the superconductor's bulk critical properties are low rather than the strength of flux pinning per se. The limiting factors for magnets can be understood with reference to the hoop stress in a magnet approximated by \cite{wilson86} 

\begin{equation} \label{hoop stress}
 \sigma_\text{hoop} = B_\text{magnet}JR_\text{magnet}, 
\end{equation}

where these are averaged properties over the magnet, and $B_\text{magnet}$ is the magnetic field, $J$ is the magnet current density, and $R_\text{magnet}$ is the radius of the magnet.

In commercial, small bore superconducting accelerator magnets such as those at CERN, the operating current density of the component superconductors is close to the critical current density of $10^9$ \si{\ampere\per\square\meter} at the operating field of 16 \si{\tesla} and the magnets are $J_\text{c}$-limited. In contrast, the relatively huge bore of our preferred fusion reactor has $B_\text{coil}$ = 12.5 \si{\tesla} and a leg-centre to leg-centre distance at the mid-plane of 9.2 \si{\meter}. At $\sigma_\text{hoop}^\text{max}$ = 660 \si{\mega\pascal} the operating current density in the TF coil winding pack is $\approx$ 2.3 $\times 10^7$ \si{\ampere\per\square\metre}, two orders of magnitude lower than that in accelerator magnets or that in the whole tape critical current density of REBCO at 12 \si{\tesla}, 4.5 \si{\kelvin} (see figure \ref{JcPlots}). It is important to distinguish whether a magnet is stress limited or $J_\text{c}$ limited. If the magnet is $J_\text{c}$ limited, improvements in $J_\text{c}$ directly increase the field the magnet can produce, whereas in a stress-limited magnet where say only a few percent of the cross-section of the coil is superconductor, improvements in $J_\text{c}$ only allow marginal increases in the steel volume content, whereas improvements in stress limits (or design) are markedly more beneficial because they increase the space for more superconductor and hence increase the operating field. If the magnet is $B_\text{c2}$-limited, increases in the superconductor upper critical field are most effective in increasing the cost-optimal field on-coil. These considerations demonstrate that in tokamaks where magnets are stress-limited and the overall current density in the cable is far from the operating $J_\text{c}$ limits of REBCO, current superconductors are far from optimised for fusion applications. REBCO magnets are stress-limited and cable design benefits most from improvements of structural material, as it is the yield stress of the material (and the design of the magnets, discussed below) that primarily determines operational limits and cost. Likewise increasing the upper critical field of Nb-Ti (e.g. via the use of a higher $B_\text{c2}$ alloy) would significantly improve its use in fusion magnets.

\subsection{Superconductors in a sea of neutron and photon radiation} \label{sea of n+p}

Superconducting magnets in a fusion environment are located in a rather special sea of neutron and photon flux, with both charged particle (ion) cascades and low energy photons continuously created in the interior (or bulk) of all the materials (cf Figure 5). Typical penetration depths for photons vary hugely as a function of energy from tens of cms at MeV to the atomic scale at meV. This means that it is difficult to artificially replicate the properties of a fusion plant flux and test a materials' performance under operation-like conditions, since high energy photons are not easily absorbed by materials, and low energy photons cannot easily penetrate material's interior. Nevertheless, the nature of the (ionised) equilibrium electronic state of a superconductor in the neutron+photon+cascade (n+p+c) sea is important since the superconducting Cooper pairs have an energy of several tens of meV ($\sim$ 3.5 k$_{B}$T$_{c}$) and it is not clear whether the pairs remain intact \cite{bardeen57}. Unfortunately theoretical considerations provide little insight, not least because we don't yet know the details of the mechanism that causes superconductivity and whether for example we should consider preformed Cooper pairs that condense, or charge carriers that both pair and condense at the same temperature. In standard superconductor lifetime experiments, superconductors are exposed to (very high) neutron flux, usually (but not always \cite{Wiesinger92}) warmed to room temperature and then cooled to have their superconducting properties measured in a radiation-free environment \cite{weber11}. However, there have been no published in-situ cryogenic measurements of the critical current density during (operational-like) n+p irradiation to confirm the superconducting properties are unaffected by the rather special sea of neutron and photon radiation produced by a fusion energy spectrum. The concern of course is that as you start turning the tritium plasma and the fusion radiation spectrum on in the tokamak, you simultaneously start turning the superconducting magnets off. Currently the community relies on at best a working assumption, perhaps guided by the uncertainty principle, that the Cooper pairs reform quickly enough for their equilibrium density to be broadly unaffected by the n+p+c fusion flux. We note that relatively low operational-level neutron flux is required for the measurements (i.e. far below those used in life-time experiments), and that data for both low temperature superconductors and high temperature superconductors are required because the electronic charge carrier densities are very different. Such measurements could be attempted using international neutron sources (the total flux at ILL is {1.5 x $10^{15}$ $n.cm^{-2}.s^{-1}$} \cite{Andersen}) or better, during a deuterium-tritium campaign \cite{Joffrin19}.

\section{\texttt{PROCESS} Power Plant Simulations}
\label{Results}

In this section we present results from our two different approaches: (a) Training: where we use Nb-Ti training magnets during the (high risk) commissioning stage (Phase 1) of a reactor that has been designed to be cost-optimised for REBCO magnets (Phase 2); (b) Upgrading: This reactor design is cost-optimised for Nb-Ti coils and later upgraded for REBCO magnets. Baseline reactors at the 100 \si{\mega\watt}$_\text{e}$, optimised for for H$_\text{98}$ = 1.2 and 500 \si{\mega\watt}$_\text{e}$ with H$_\text{98}$ = 1.6 were generated for both approaches. For each of these baseline reactors the geometry was fixed and the superconducting materials used in the TF and CS coils were replaced with either REBCO, commercial Nb-Ti or quaternary Nb-Ti.  By ``reactor geometry'', we mean the reactor's physical build (e.g. the location of the coils, thickness of blanket, location of the first wall, etc.) - the plasma shape was allowed to vary. In both approaches, the magnet systems from the first phase were replaced by magnets where the combined sizes of the CS and TF coils changed by less than 1\% (equivalent to re-optimising the cable dimensions and construction, but keeping the reactor radius and height fixed). In all cases the second phase reactors had magnets re-optimised for maximum net electricity yield in order to maximise the swapped magnets' performance. For each of the six baseline tokamaks shown in Tables (7 - 12), we have also calculated the effect a reduction in H$_\text{98}$-factor to a value of unity would have. These data show the drop in fusion power and net electricity generation that would occur were plasma quality not to meet the higher values hoped for. The six baseline reactors considered in this paper (including the preferred reactor in bold) are compared to other world-wide tokamak designs in table \ref{ReactorComparison}..

For the training approach, the capital costs were calculated by adding the cost of the full-power reactor to the cost of the training magnets alone. We have assumed that other plant components can simply operate with reduced capacity. For upgrading approach, the capital costs were calculated as the cost of the respective baseline reactor combined with the cost of the new magnets and costs associated with increasing the scale of additional plant components (enhanced generator capacity, heat transport, fuel handling etc.). In all cases as noted above, we have not added cost associated with making the magnets remountable or the design and operational costs associated with robotic handling. During the commissioning of the reactor, the full-power REBCO coils must of course be tested and commissioned themselves. Indeed the reactor will have to itself be recommissioned at full power.The training coils will allow allow the operational team to iron out the majority of user error and manufacturing error related disruptions and unexpected events which could destroy the brittle REBCO magnets, before they are installed. Plasma will be generated, power plant systems will be able to be tested etc. – simply not at full capacity, but close enough to full capacity to discover and significantly reduce the risk of magnet-destruction-capable events. Experience has shown that the majority of disruptions occur at the beginning of reactor life \cite{devries11} - using training coils therefore puts the brittle full-power magnets at much less risk.

\subsection{REBCO tokamaks with training magnets}
\label{results-REBCO}

Table \ref{TrainingH=1.2} shows that for our preferred reactor, the use of commercial Nb-Ti TF and CS training magnets during the training phase reduces fusion power to $\approx$ 60 \si{\mega\watt} from 860 \si{\mega\watt}, and results in an electricity deficit of $\approx$ -180 \si{\mega\watt}$_\text{e}$. Despite the less energetic plasma and the lower peak fields on TF coils (70 \%) and CS coil (66 \%) that commercial Nb-Ti training magnets would generate, they would nevertheless allow rather thorough machine testing during the plant commissioning phase. The table shows that the final total cost for the training magnets and the final preferred tokamak together is \$ 4.54 Bn (1990 US) equivalent to about \$ 9.75 Bn (2021 US) - along the lines of \$ 20 Bn estimated for a DEMO reactor \cite{Cardozo16}. Interestingly, quaternary Nb-Ti TF training magnets are almost able to match the field of the full-power REBCO TF coils (93 \%) when REBCO is used for the CS coil. These large percentages are because the high $B_\text{c2}$ quaternary Nb-Ti coils are able to produce a large fraction of the stress limited cost-optimal field and provide a prima facie case for the fusion community to develop as a priority, new high $B_\text{c2}$ ductile low temperature superconductors specifically for fusion, but with $J_\text{c}$ values that by the standards of other applications are (undemandingly) low. At H$_\text{98}$ = 1.6 (Table \ref{TrainingH=1.6}) the lower plasma current reduces the peak field on the REBCO baseline CS coil, so commercial Nb-Ti CS coils produce 77 \% of the CS coil peak field and the quaternary Nb-Ti coils produces 87 \%. Similarly, higher H-factor means the CS training coils also do not have to produce as high a magnetic flux. The commercial Nb-Ti CS and PF system delivers only 105 \si{\weber} compared to the 261 \si{\weber} of the training coils for the preferred reactor (note that the plasma current fractions of the respective baseline reactors are however approximately equal at 62 \% at H$_\text{98}$ = 1.6 and 66 \% at H$_\text{98}$ = 1.2). Given the peak field on the CS remains 9.2 \si{\tesla} in both cases and the number of turns falls by only 10 \%, the lower flux requirement allows a larger proportion of the available CS-TF space to be occupied by the TF coils and a 32 \% larger TF conductor cross-section as a result - allowing for the production of a larger toroidal field. The quaternary Nb-Ti case is similar.

\begin{figure}
    \centerline{
    \includegraphics[height=9cm,keepaspectratio]{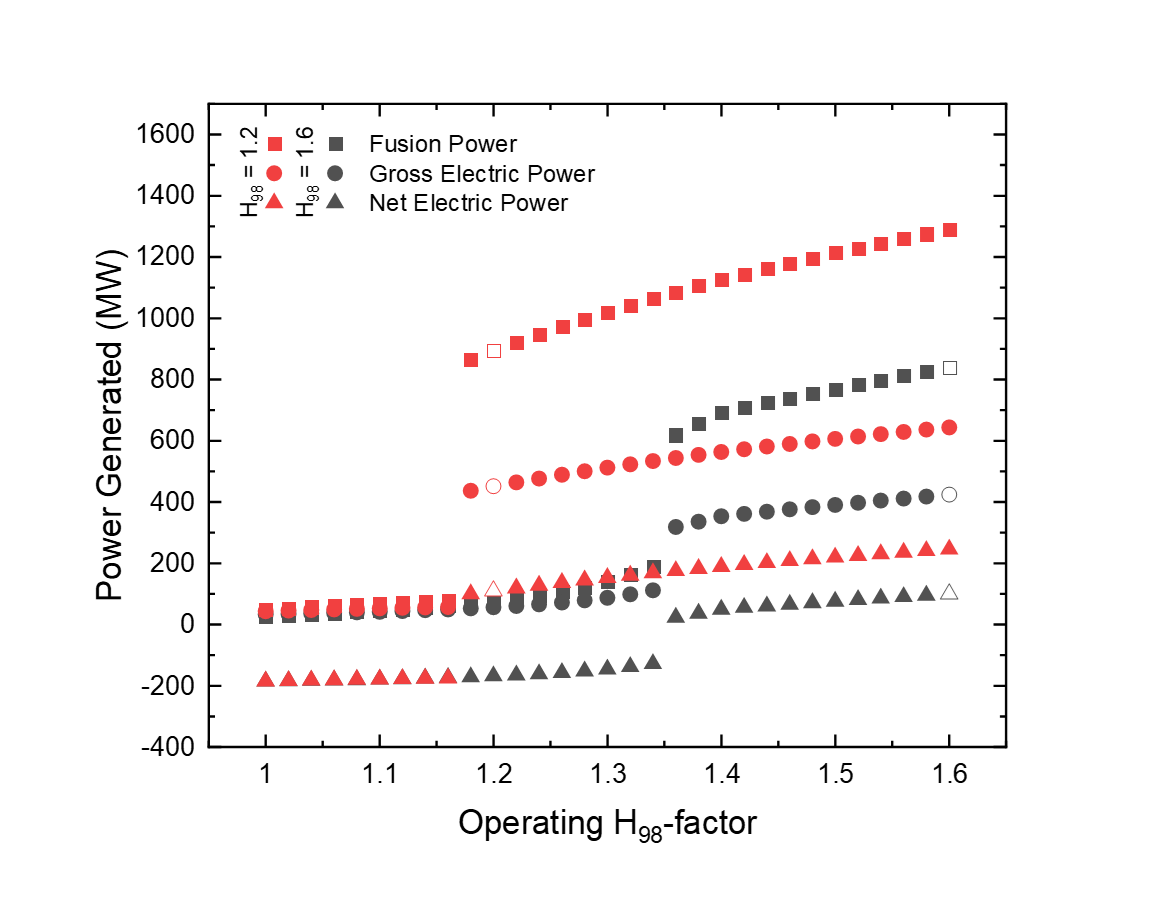}}
    \caption{\label{PowervsHfactor}Fusion power, gross electric power and net electric power as a function of operating H$_\text{98}$-factor for two reactors with REBCO CS and TF coils and Nb-Ti PF coils. The reactors were optimised for minimum capital cost (open data points) at H$_\text{98}$-factors of H$_\text{98}$ = 1.6 (black) and H$_\text{98}$ = 1.2 (red) and produce 100 \si{\mega\watt}$_\text{e}$. Reactors at higher or lower than expected H$_\text{98}$-factors were optimised to produce maximum net electricity (solid data points). The discontinuities in fusion power that occur between high and low H$_\text{98}$ factors are due to a loss of energy confinement at H$_\text{98}$ factors that are too low.}
\end{figure}

The effect of a range of different H$_\text{98}$-factors (H$_\text{98}$ = 1.2 and H$_\text{98}$ = 1.6) for two REBCO 100 \si{\mega\watt}$_\text{e}$ reactors are shown in figure \ref{PowervsHfactor}. For increases in H$_\text{98}$ factor above design expectations the fusion power is not greatly increased. However, if in practice the H$_\text{98}$ factor achieved is below the design specification, the plasma loses energy faster than it is supplied, energy confinement is lost \cite{jakobs14} and the plasma burn cannot be maintained for the required 2 hours. As a result the fusion power collapses, resulting in negative net electricity production. It is therefore important that tokamak power plant designs are conservative with regard to H$_\text{98}$-factor, since the consequences of unexpectedly poor performance are quite severe.

If we consider building a very large 500 \si{\mega\watt}$_\text{e}$ REBCO baseline design, the reactor with commercial Nb-Ti TF and CS training magnets is able to generate 78 \% field on TF (Table \ref{Training500}). This larger percentage is because the TF coils of the 500 \si{\mega\watt}$_\text{e}$ reactor have larger radii (11.1 \si{\meter} from leg centre to leg centre at the mid-plane compared to the 9.2 \si{\meter} of the 100 \si{\mega\watt}$_\text{e}$ preferred reactor) which reduces the optimal field on TF for the baseline design due to the larger stresses (equation \ref{hoop stress}). The larger plasma current requirement for the larger fusion power means that the poloidal field coils generate a larger proportion of the magnetic flux (it is more cost-beneficial to increase the output of the Nb-Ti poloidal field coils, than it is to greatly increase the size of the REBCO CS and overall reactor volume as a result). Here, the PF system delivers 51 \% of the total magnetic flux, compared to 43 \% of the flux in the 100 \si{\mega\watt}$_\text{e}$ REBCO reactor. The flux requirement when the commercial Nb-Ti training coils are used drops from 446 \si{\weber} to 321 \si{\weber}, so now the (unchanged) PF system delivers 66 \% of the flux, reducing the flux demand on the CS, reducing its necessary size and allowing for larger TF coil thickness. The larger cross-section of the TF coils of the commercial Nb-Ti training coils for the 500 \si{\mega\watt}$_\text{e}$ reactor leads to a 24 \% larger conductor cross-section (in comparison to the 100 \si{\mega\watt}$_\text{e}$ reactor case) and the generation of an additional $\approx$ 0.5 \si{\tesla} on-coil. Commensurately, a full quaternary Nb-Ti set of training coil can generate a TF coil field of nearly 97 \% of that of REBCO with a field on coil of 11.5 \si{\tesla}.

All three baseline reactors in Tables (\ref{TrainingH=1.2} - \ref{Training500}) operate with TF and CS coils at the stress limit of 660 \si{\mega\pascal}. Taking the case of the 100 \si{\mega\watt}$_\text{e}$ preferred reactor, when only commercial Nb-Ti TF training coils are used, operation close to  upper critical field demands a larger superconducting fraction in the cable and reduces TF coil steel fraction from 52.8 \% to 38.5 \%. Operation below the designed-for field reduces stress on the TF by 270 \si{\mega\pascal} and CS by 210 \si{\mega\pascal}. When both TF and CS training coils are used the steel fraction must also decrease (from 80.4 \% to 42 \%) in order to maximise the CS superconductor fraction and magnetic flux the coil can generate. The peak stresses rise closer to the 660 \si{\mega\pascal} limit as the interplay between the TF and CS coil thicknesses allows for better optimisation of the coil steel fractions.

\subsection{Nb-Ti Tokamaks with Upgrade Coils}

Table \ref{UpgradeH=1.2} shows that upgrading tokamaks is a significantly more expensive approach than the training approach (cf 6470 M\$ in Table \ref{UpgradeH=1.2} compared to 4540 M\$ in Table \ref{TrainingH=1.2}). Although swapping the CS and TF coils for REBCO, or the CS coil for REBCO and the TF coil for quaternary Nb-Ti produces more electricity (i.e. $\approx$ 280 \si{\mega\watt}$_\text{e}$) and swapping all coils for quaternary Nb-Ti yields more electricity (i.e. $\approx$ 230 \si{\mega\watt}$_\text{e}$), we feel these increases do not significantly better de-risk fusion energy production for commercialisation (considered in detail below). In the former case, the CS and TF coils are stress limited, so although in principle the REBCO upgrade coils could produce higher fields on the plasma and in the CS coil, they are prevented from doing so by the thickness of the steel support required to resist the greater magnetic forces in the limited space available. In the latter case, the limiting factor is the critical current density of the quaternary Nb-Ti cable, in the CS coil. Just swapping the centre solenoid alone for REBCO, while keeping the original commercial Nb-Ti TF coils offers no benefit, as the TF coils in the baseline design are already $B_\text{c2}$ limited. The H$_\text{98}$ = 1.6 commercial Nb-Ti case, shown in table \ref{UpgradeH=1.6}, is very similar to the H$_\text{98}$ = 1.2  case. The larger fields produced by the upgraded magnets are due to the higher H$_\text{98}$-factor reactor's smaller major radius, meaning that the stresses on the TF coils are consequently lower for a given field-on-coil. 

For the larger 500 \si{\mega\watt}$_\text{e}$ tokamak (Table \ref{Upgrade500}), upgrading the reactor with a REBCO CS coil and either REBCO or quaternary Nb-Ti TF coils results in an increase in net electricity output of $\approx$ 36\%. The smaller percentage increase in this larger machine is due to the more stringent stress limits in the larger radius coils. Indeed, a fully quaternary Nb-Ti upgraded reactor would only generate 8 \% more electricity as the current in the CS coil is limited by the critical current density of the conductor. 

The cost-optimised upgraded tokamaks again operate with 660 \si{\mega\pascal} stresses on the TF and CS coils in order to minimise the amount of steel support used. Focusing on the 100 \si{\mega\watt}$_\text{e}$, H$_\text{98}$ = 1.2 reactor, when REBCO TF and CS coils are used, the peak fields on coil increase by 9 \% and 63 \%, the current densities increase by 1 \% and 14 \% and the coil radii decrease by 0 \% and by 6 \% respectively. The resulting change in stress (equation \ref{hoop stress}) necessitates increases in the TF and CS coil steel fractions by 2.3 \% and 27.6 \%. The same is true of quaternary Nb-Ti upgrade coils, though the increases in steel fractions are more modest due to the lower increases in field. 

\subsection{Spherical Tokamak Power Plants}

Spherical tokamaks have some significant advantages and disadvantages over conventional aspect ratio tokamaks and are being considered for pilot fusion power plants \cite{STEP20, sykes18}. Spherical tokamaks operate at higher beta (of up to 40 \% \cite{robinson99}) and at higher safety factors (e.g. $q_{95} = 8.9$ in FNSF \cite{menard16}) than conventional reactors, meaning that that their plasmas are inherently more stable (and disruptions less likely) for a given field on plasma axis \cite{robinson99,peng00,friedberg07}. Spherical tokamak pilot plants have proposed designs that are compact, with major radii $< 3.5$ \si{\meter} \cite{sykes18,wilson04,najmabadi03,menard16} (in principle reducing construction costs and time \cite{windridge19}) whilst producing a plasma fusion gain $\approx$ 30 \cite{sykes18,costley19}. 

Spherical tokamaks’ compact size however also increases average neutron wall loading, above 3.5 \si{\mega\watt\per\square\meter} in some proposed pilot plant designs \cite{wilson04,najmabadi03}, to more than three times that of EU-DEMO \cite{federici18}. This very high flux necessitates thick radiation shielding (or breeding blanket) of $\approx$ 60 \si{\centi\meter} for the central column \cite{menard16} (reducing the field on coil for a fixed reactor major radius), or frequent remote replacement of the central column magnets (on the order of every 3 years \cite{najmabadi03}). The small size also increases the power through the separatrix (above 30 \si{\mega\watt\per\meter} in some designs \cite{wilson04,najmabadi03}) necessitating the use of advanced divertor configurations \cite{fishpool13} which would either make use of sacrificial, resistive, inboard coils (that are part of a higher order reactor design than that discussed in this work) or require heavily distorted TF coil architectures \cite{ambrosino19,reimerdes20}. Large tokamak design studies of ITER-type plants have shown the cost of electricity is lower for large tokamaks, scaling proportionally to the electric power of the tokamak raised to the power -0.59 (i.e. increasing the electric power produced by a factor 10 reduces the cost of electricity by about 4) \cite{lee15} and ultimately means that large fusion power plants are likely to produce the cheapest electricity. Nevertheless compact spherical tokamaks may offer the opportunity for lower capital costs to demonstrate fusion energy is commercially viable - it is beyond the scope of this paper on ITER-like aspect ratio machines to assess to what degree these reductions are offset by the specific technical challenges of spherical tokamaks, and hence how effectively they provide a short-cut to demonstrating commercial fusion energy is cost-effective.

\subsection{Aluminium/Copper Tokamak Power Plants}
\label{resistive_tokamaks}

Copper \cite{saleh14,troxell89} and aluminium \cite{flores18} have the best combination of high strength and high electrical conductivity to have been the choices for for magnets in experimental fusion reactors. Resistive magnets for tokamaks are typically operated at 300 – 400 \si{\kelvin} and do not suffer the same cut-off in current carrying capacity with neutron fluence as superconductors, meaning that shielding requirements are reduced and reactors can (in principle) be made more compact. However, the (magneto-)resistivity of these materials is unchanged for many decades and can be contrasted with developments in superconductivity where large scale projects such as ITER and CERN continue to drive improvements in materials with higher current density, and one can expect commercial fusion to drive the development of the more neutron tolerant materials (eg high-field alloys). Resistive tokamaks have indisputably helped drive our experimental understanding of, and encouraged new designs for, high-field fusion plasma physics. For example, ARIES-ST \cite{najmabadi03} which was designed with only a 20 \si{\centi\meter} ferritic steel centre-post shield leading to a predicted nuclear heating of 164 \si{\mega\watt} and total magnet system losses of $\approx 730$ \si{\mega\watt}$_\text{e}$, an order of magnitude more than the cryoplant power and magnet system losses for the superconducting reactors in table \ref{Power_Breakdown}. 

Proposed resistive reactors however have very low plasma burn times (e.g. FIRE \cite{meade02} (a prototype reactor) with a 20 \si{\second} plasma burn and $\approx$ 3 h repetition time) or rely on non-inductive start-up mechanisms and assume large $H_\text{98} \gtrsim $ 1.5 (e.g. ARIES-ST \cite{najmabadi03} and STPP \cite{wilson04}). Even the small size of proposed resistive tokamaks would not obviously reduce their capital cost compared to superconducting reactors. Both ARIES-ST’s cost of $\approx$ 4200 M\$ (1990 \$) and our \texttt{PROCESS} generated capital costs for a 100 \si{\mega\watt}$_\text{e}$ STPP-like reactor (with minimised capital costs) of 5200 M\$ shown in \ref{CopperPilot} are similar to the cost of the $H_\text{98}$ = 1.2 preferred reactor, and in fact larger than the $H_\text{98}$ = 1.6 REBCO reactor in this work. The reduced costs due to smaller size are counterbalanced by cost increases in coil bussing and power conditioning. It has been argued that high temperature superconductors are an essential technology that will enable commercial fusion power \cite{whyte16}. We broadly concur that the scarce resources for commercialisation of fusion are best focussed on de-risking and up-skilling in commercialising the unprecendently large superconducting magnets required for fusion, rather than resistive ones.

Using a similar validation process to the one above, we first confirm the reliability of our \texttt{PROCESS} calculations by comparing them with some simple benchmark calculations and with JET experimental results. Our approach was to compare the maximum current per resistive magnet turn to the maximum current per superconducting magnet turn in a tokamak with space allocated to the magnets equal to that in our preferred tokamak. Note that a turn consists of the cable and the insulation. The cable has a structural, conducting and cooling channel components. In our \texttt{PROCESS} simulations the insulation is 1.5 \si{\milli\meter} thick. Making explicit our use of Ampere's law \cite{hampmax18}, we can write:
\begin{equation}\label{Bcalc}
I^\text{resistive} _\text{turn} / I^\text{preferred}_\text{turn} = B_\text{T}^\text{resistive} / B_\text{T}^\text{preferred} 
\end{equation} 
where $I^\text{resistive} _\text{turn}$ and $I^\text{preferred}_\text{turn}$ are the coils’ currents per turn on the resistive reactor and the preferred choice reactor and $B_\text{plasma}^\text{preferred}$ and $B_\text{plasma}^\text{resistive}$ are the magnetic fields on plasma axis of the preferred choice reactor and resistive reactor respectively. 
The maximum current per resistive magnet turn was calculated using 
\begin{equation}
Q_\text{cooling}^\text{max}=(I^\text{resistive}_\text{turn})^2\frac{\varrho_n L_\text{turn}}{(A_\text{turn} - \pi(D_h/2)^2)}
\end{equation}
where $\varrho_n$ is the conductor resistivity, $L_\text{turn}$ is the length of each turn and $A_\text{turn}$ is the cross-section of each turn. The maximum cooling power per turn  $Q_\text{cooling}^\text{max}$ is given by
\begin{equation}\label{cool_max}
Q_\text{cooling}^\text{max} = \dot{m} c_p \Delta T_{turn}
\end{equation}
where $\dot{m}$ is calculated from equation \ref{eq6} noting that for $\dot{m} \approx $ 0.1 \si{\kilo\gram\per\second} $f_d \approx$ $0.0231/ V_{coolant}^{0.742}$. \par
In each of JET's 32 TF coils \cite{bertolini00,mlynar07,JET82}, there are 24 turns of length 15 \si{\meter}, and average turn cross section of $\approx$ 32 \si{\square\centi\meter}. Each turn has its own cooling channel with a cooling channel hydraulic diameter of $D_h$ = 1.5 \si{\centi\meter}, an inlet over-pressure is 5 \si{\bar} and $\Delta P$ = 0.5 \si{\bar}. The temperature of the coolant is $T_\text{coolant}^\text{inlet}$ = 293 \si{\kelvin} and $\Delta T_{turn}$ = 75 \si{\kelvin}. We have set the average copper magnetoresistivity over the temperature range (and typical field $\approx$ 6 $\si{\tesla}$) to be 2.20 $\times 10^{-8}$ \si{\ohm\meter} (c.f. Table \ref{Cryoproperties} \cite{matula79,simon92}) and used the well-known properties of water for the coolant (rather than the anticorrosion fluid Galden HT55 used in practice). The simple benchmarking equations above yield a current per turn of 63.4 \si{\kilo\ampere}, very close to the JET current per turn of 66 \si{\kilo\ampere} and resistive losses for the TF magnet system of 497 \si{\mega\watt}, close to the JET power requirements of 700 - 1375 \si{\mega\watt} \cite{JET82, JET_power_supply}.

Turning to use the geometry of our preferred choice reactor, there are 18 TF coils each with 100 turns each of length  38.1 \si{\meter} and a cross-section for each cable of 39.6 \si{\square\centi\meter}. Following JET, we consider resistive copper magnets operating at room temperature where each turn has its own cooling channel. The inlet pressure was set to that of JET with 5 \si{\bar} and $\Delta P$ = 1 \si{\bar}, as well as the temperature of the coolant, $T_\text{coolant}^\text{inlet}$ = 293 \si{\kelvin} and $\Delta T_{turn}$ = 75 \si{\kelvin}. The cooling channel hydraulic diameter was optimised and found to be at 56 \% of the cross-section (i.e. $D_h$ = 5.3 \si{\centi\meter}). Equations (\ref{Bcalc} - \ref{cool_max}) then yield a maximum current per turn of 57.1 \si{\kilo\ampere} compared to the preferred reactor which has 100 \si{\kilo\ampere} per turn. Not surprisingly for resistive magnets, we find a huge power consumption of 3 \si{\giga\watt}. We now consider whether running resistive magnets at cryogenic temperatures is more viable and calculate the performance of cryogenically cooled resistive aluminium magnets cooled to 65 K using liquid nitrogen as the coolant, and then at 20 \si{\kelvin} using supercritical helium. Aluminium and copper have similar room temperature resistivity ($\approx$ 2.7 $\times 10^{-8}$ \si{\ohm\meter} and $\approx$ 1.7 $\times 10^{-8}$ \si{\ohm\meter} respectively), but it is cheaper to make high-purity high-strength aluminium than copper, so aluminium is generally preferred at low temperatures for cyocooled resistive magnets. Ensuring the nitrogen doesn't solidify or become gaseous requires setting $T_\text{coolant}^\text{inlet}$ = 65 \si{\kelvin} and $\Delta T$ = 15 \si{\kelvin}. Averaging over the temperature range, at 6 \si{\tesla} and RRR = 10000, Al has a resistance of $\approx$ 3 $\times 10^{-9}$ \si{\ohm\meter} \cite{egan90,fickett72}. Using the benchmarking calculations we find the current in each turn is only 35.6 \si{\kilo\ampere} which is lower than the 100 \si{\kilo\ampere} in the preferred superconducting reactor. Of greater concern commercially (also found below for 20 \si{\kelvin}) is that the required resistive heating is 148 \si{\mega\watt} which even with ideal Carnot losses require a cryocooler electric power of 683 \si{\mega\watt}$_\text{e}$. Turning to 20 \si{\kelvin} operation with $\Delta T$ = 20 \si{\kelvin} and using supercritical helium: averaging over the temperature range, at 6 \si{\tesla} and RRR = 10000, Al has a resistance of $\approx 10^{-10}$ \si{\ohm\meter} \cite{egan90,fickett72}. Under these conditions, the current in each turn is higher than the superconductor by about 50 \% (were it not to be stress limited), however the resistive losses of the Al magnet system would be $\approx$ 90 \si{\mega\watt} at 20 \si{\kelvin} equivalent to a huge cryocooler electric power 1.35 \si{\giga\watt}$_\text{e}$. Our calculations show why superconductivity is a disruptive technology for fusion: resistive magnets in a fusion power plant would consume most of the  power (and sometimes more) than the plant would produce. At room temperature, huge levels of power are needed to drive the magnets themselves. At cryogenic temperatures, equally huge levels of power would be required to drive the cryoplant.

Finally we note that plasma control is more demanding with resistive magnets compared to superconducting ones. When the current changes in a resistive magnet, along with the magnetic field changing, the temperature and the size of the magnets change because of the thermal expansion of the various components. In the superconducting case, the current does not substantially heat the magnets so the differential thermal properties of the component parts of the tokamak play little role. We have not included any calculations for the cost of robotic replacement of resistive parts, which will almost certainly be cheaper than removing and installing brittle superconducting components. However, we have not proceeded any further with calculations for resistive magnets given their huge power demands, and that we feel the magnetoresistance of copper and aluminium is well enough understood that no significant reduction in the magneto-resistivity is likely.  Our calculations, table \ref{CopperPilot}, show that even if we could find some way for the plasma to perform well beyond current expectations at $H_\text{98}$ = 1.6, we might get a little electricity. In the context of identifying the remaining challenges of making fusion energy commercial, engineering large robotically-remountable superconducting magnets is best explicitly included for clarity of purpose \cite{UKFusionStrat}. Unfortunately resistive magnets are a very mature technology and history teaches us what inevitably happens to technologies that can't evolve, if they compete commercially with a continuously improving disruptive technology - superconductivity. Hence, we set aside considering large resistive magnets in the tokamak for the rest of the paper.

\section{Future Technological Developments}\label{Future developments}

Magnet technology is a rapidly evolving field. In this section we address whether the inevitable improvements in technology that are on the horizon are likely to bring significant cost reductions.  We discuss how increases in structural steel yield stress (which we use as a proxy for both improved strength of materials and improved magnet design support architectures) and how reductions in REBCO cost would affect our preferred reactor design and capital cost. Then we discuss the utility of producing some new fusion-specific high $B_\text{c2}$ superconductors, some disruptive designs and finally some general comments about future costs of tokamaks.

\subsection{Novel Support Architectures and Strengthened Steels}\label{strengthened_steels} 

\begin{figure}
    \centerline{
    \includegraphics[height=14cm,keepaspectratio]{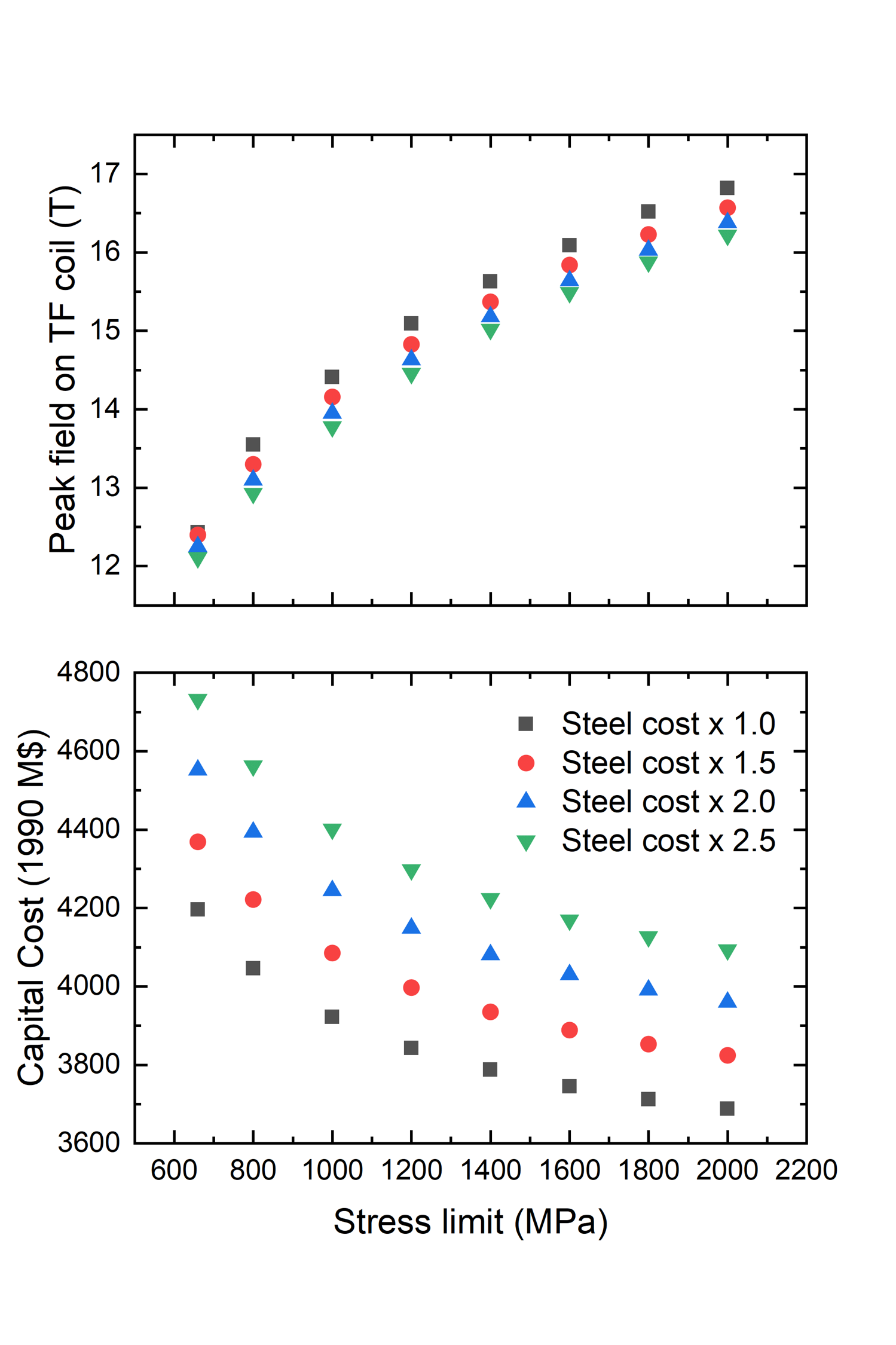}}
    \caption{\label{SteelYS} (a) Change in cost-optimal field on coil and (b) capital cost of the preferred reactor (a cost-optimised H$_\text{98}$ = 1.2, 100 \si{\mega\watt} net electricity tokamak with REBCO toroidal field and central solenoid coils) as a function of the allowable maximum of the shear stresses (as used for the Tresca yield criterion) on the central solenoid and inboard toroidal field coil mid-planes. Different data sets correspond to different costs of steel components (standard, 1.5 $\times$ standard etc.), representative of either more expensive steels or larger steel volumes.}
\end{figure}

Although it seems unlikely that the strength of (steel) materials will very siginficantly be increased (eg that the Tresca yield criterion (used in PROCESS) will be increased much beyond 660 \si{\mega\pascal} - 2/3 yield strength of 316LN steel \cite{hamada07}), improved designs of the external support structures such as those proposed for the ARC reactor \cite{sorbom15}, do reduce the stresses on the TF coils. For example, the ARC support rings at the top and bottom of the TF coils resist both toroidal and vertical forces, and have been shown to increase very markedly the possible fields on plasma. In figure \ref{SteelYS}, data are included for increasing the TF and CS coils' operational stress limits for various inflated costs of complete structural steel components; representative of increased cost of enhanced steels or increased volumes of steel used in advanced support architectures. Also shown are data for decreased cost versus increased stress limits, which are a proxy for improvements in design rather than the steel itself. Increasing the TF and CS coils' operational stress limits to 1000 \si{\mega\pascal} in our preferred choice reactor increases the cost-optimal field on coil to 14.4 T ($B_\text{plasma} = 6.3 $ T) and reduces the capital cost to 3920 M\$ (assuming steel costs do not change). With larger allowed stresses the field increases further, plateauing by $\approx$ 2000 \si{\mega\pascal} with a field on coil of 16.8 T ($B_\text{plasma} = 7.7$ T) and capital cost of 3690 M\$ (due to the $P_\text{sep}/R_\text{major}$ = 20 \si{\mega\watt}/m$^{-1}$ limit). These magnetic fields approach those of the SPARC and ARC tokamaks (cf Table 1) that include more advanced designs of the TF coils. However, the trade-off between the reduction in reactor capital cost associated with a reduction in reactor volume enabled by larger toroidal fields, and the additional cost of the support structure required to reach these fields  means, as shown in figure \ref{SteelYS}, that the capital cost reductions are modest.

\subsection{Fusion-Specific Superconducting Materials}

 The superconducting strands and tapes currently used to develop cables for fusion tokamaks were optimised for other applications. The Nb47\%wt.Ti alloy used in ITER was developed for the MRI market to maximise $J_\text{c}$ between 4-6 \si{\tesla}; the Nb$_3$Sn strands were developed for particle accelerator magnets to maximise $J_\text{c}$ between 8-10 \si{\tesla}; and the aim of the REBCO industry has focussed on developing higher $J_\text{c}$ tapes (at reduced cost) for power applications and ultra high field magnets. REBCO cables for tokamaks are currently being optimised for immediate use - CORC \cite{weiss17}, slotted core cables \cite{hartwig20,celentano14}, twisted-stack cables \cite{bykovsky16,wolf16}) - but commercial fusion will drive the development of new REBCO conductors. As we have seen in section \ref{Results}, the huge fusion magnets in our optimised REBCO design include magnet support structures that are stress limited. Indeed, the critical current density of the REBCO tapes is already two orders of magnitude larger than current density of the winding pack, which opens the possiblity of developing cheaper, lower $J_\text{c}$ tapes (that are already available) and that can also help mitigate other important issues such as quench mitigation and brittle fracture. We note for example that were commercial quaternary Nb-Ti training magnets available, they would produce a 10 - 20 \% larger field fraction than conventional Nb-Ti - an increase driven solely by the alloy's larger upper critical field. 

It is a concern for rapid commercialisation of fusion energy that commercial HTS manufacturing capability is modest. Nevertheless, fusion power plants represent a multi-billion-dollar market for high temperature superconductors, and ITER demonstrates a precedent for rapid growth in superconductor manufacturing capacity when required. ITER required $\approx$ 600 tons Nb$_3$Sn strands \cite{devred14} which led to a rapid increase in annual global production from $<$ 2 ton/year in the early 1990s to 100 tons/year today \cite{uglietti19}. Equally, REBCO tapes have seen a substantial decrease in production cost in the two decades \cite{cooley16} and it is likely that this will continue as global demand increases - note that doubling superconductor $J_\text{c}$ (for the same manufacturing cost) has the same effect as halving cost in \$/\si{\kilo\ampere\meter}, see equation \ref{cost_equation}. Remaining focused on our preferred choice reactor design, if we reduced the REBCO cost from 30 \$/\si{\kilo\ampere\meter} (6 \si{\tesla}, 4.2 \si{\kelvin}) to make it almost free (0.025 \$/\si{\kilo\ampere\meter} (6 \si{\tesla}, 4.2 \si{\kelvin}), the resulting capital cost minimised reactor has an increased toroidal field of 12.7 \si{\tesla} and a capital cost reduced from 4230 M\$ to 4000 M\$. It is noteworthy that these very modest changes in cost are less than those obtained by increasing the steel stress limits. We conclude that a reduction in REBCO cost beyond those presented in \cite{cooley16} would not offer much benefit in overall reactor cost as the cable cost and design-limiting factors are dominated by non-superconducting components.

\subsection{Driving down Costs}

Paymasters inevitably ask whether there may be any developments in the future that are likely to substantially reduce costs. For fusion technology this will be informed by the commercial research that will be completed using the prototype, and that may enable some additional reductions in costs or improvements in operation. 

In this context, the exciting very high field values for the ARC and SPARC tokamaks (noted in Table \ref{ReactorComparison}) using high temperature superconductors, can be seen as both derisking a fully integrated tokamak with high power density and gain, while also enabling a search for more stable plasma operation at the highest fields available. However, small tokamaks bring high power fluxes and make radiation-hard robotic handling more challenging, and remind us that there may be more than one successful approach to commercial fusion technology.

If we use our minimum cost approach using \texttt{PROCESS} to consider a tokamak with a TBR reduced from 1.1 to 0.9, it produces a capital cost reduction of $\approx$ 24 \% , not just from the reduction in the cost of the blanket (which makes up $\approx$ 12 \% of the preferred reactor direct cost, see table \ref{Cost_Breakdown}) but also by reducing the reactor volume (by reducing the inboard and outboard blanket modules’ thicknesses by 64 \%) by 28 \%.  Given that magnetic technology is the primary driver for the tokamak size and hence cost, we again use the the maximum yield stresses as a proxy for better design of/stress management in the TF and CS coils and set it (unrealistically high) to 1000 \si{\mega\pascal}. We also set the REBCO cost unrealistically low to  0.025 \$/\si{\kilo\ampere\meter}, the H-factor set to $H_{98}$ = 1.8 and the limit on the power through the separatrix was increased to $P_\text{sep}/R =$ 25 \si{\mega\watt} as further proxies for reasonable improvements in design and materials improvements. These (unrealistic nay extreme) choices provide a means to find just how sensitive capital cost can be to future developments. All other constraints from our preferred choice reactor were retained. Under these conditions, the optimised reactor has a field on plasma of 5.7 \si{\tesla}, field on TF coil of 14.0 \si{\tesla}, peak field on CS coil of 12.4 \si{\tesla}, plasma current of 9.7 \si{\mega\ampere}, major radius of 5.37 \si{\meter}, aspect ratio of 3.36 and a capital cost of 3400 M\$. This leads to a cost only 19.5 \% lower than our preferred reactor.  

To drive costs down even further and pursue the cost of a cheapest possible tokamak demonstrator, we combined these approaches to consider a machine that only produced enough electricity to break-even, but would otherwise demonstrate all the core magnet technologies, capabilities of the blanket and successful coupling to electricity generating turbines. Keeping the same low REBCO costs, allowable maximum of the shear stresses and burn time as in our preferred reactor, with a TBR = 0.9, such a tokamak would need to produce $\approx$ 560 \si{\mega\watt} fusion power and would have a capital cost of 3170 M\$, about 31 \% lower compared to our preferred reactor. The reduced blanket size would allow for a more compact $R_\text{major}$ = 6.2 \si{\meter}, operating with a field on plasma of 6.1 \si{\tesla} and field on TF coil of 12.5 \si{\tesla}. These calculations show the limitations of the further cost reductions that are possible.  We suggest that finalising the best tokamak to build would include consideration of the possible potential developments in fusion-focused high B$_{c2}$ alloys, as well as whether or not large fusion power plants are inevitably the commercial endgame. We conclude that if we require retaining the CSC, we can reduce the cost by up to one third if we reduce the performance and fast-track design and material improvements. Naturally, investors (who have so far invested more than \$ 2.4 B) \cite{Ball21} would want to ensure that such a reduction in cost and performance nevertheless still de-risks the commercial build. 

\section{Discussion}
\label{Discussion}

The scientific evidence for the climate emergency \cite{IPCC2021} and the damage it is causing is now sufficiently clear \cite{lustgarten20,mcmichael11} that there is a global commitment to zero carbon \cite{ParisAgreement}. For example, the UK Government has banned the sale of new cars powered solely by petrol or diesel from 2030 \cite{johnson20} and a legally binding commitment to ``at least" zero carbon by 2050 \cite{shepheard20}.
 
Market-ready, intermittent solar and wind resources are the natural place to start. 
However the decarbonisation commitments to replace the carbon fuels used both in power stations and in transportation in the UK, will require several orders of magnitude more green electricity than is currently planned for. Unfortunately, unlike other low-CO$_2$ energy sources, although a relatively small power plant can de-risk the holistic integration and operation of the key-technologies for commercial fusion, cost of electricity will eventually drive commercial fusion power plants to be large \cite{lee15} (as is the standard for fission plants; Hinkley C will produce $>$ 3 \si{\giga\watt}$_\text{e}$ \cite{HinkleyC}. To what degree fusion power will be used to provide base-load electricity directly \cite{fedkin20}, rather than produce synthetic fuel for say aviation \cite{henning}, or hydrogen \cite{Ward09}, or conversion of carbon dioxide back into carbon black \cite{NECOC20}) or reduce methane \cite{COP26} are open questions that will depend on the particular commercial realities of global warming when commercial fusion energy arrives, and whether we will need it to reverse global warming.
\par
Fusion energy is clearly a huge-risk huge-return investment that only a relatively small group of wealthy Governments \cite{STEP20,NIST21,DEMO20,zhuang19} or philanthropic billionaires \cite{polio21} can lead. Much more work is needed to bring the scientists, policy makers, investors, defence interests, and public together to de-risk fusion power and make it commercial as quickly as possible.
Even the excellent texts that deal with the science of renewable energy \cite{mackay08} or the economic opportunity (or green premium) it presents \cite{carney21}, hardly mention fusion at all. Fusion technology is quite different from other renewable technologies in that the scale of investment required to make meaningful commercial advances is much larger.  
Of course there have been huge projects in the past, including space travel (i.e. putting the first man on the moon \cite{kennedy62} and the development of fission power). Those projects had straightforward aims and well-defined competitive environments, but crucially, did not have to broadly operate in the free-market while being commercialised. Commercialising fusion is far more complex. The green changes that are underway were not driven by the free-market, but by an unprecedented alignment of public social awareness and enlightened policy. However in order for the commercialisation of fusion energy to be a success, it must roll-out across the developed world on a huge scale. It therefore needs the resources and skills of the markets with commensurate and proper financial returns for investors.

The approach for designing, financing and building the last fusion tokamak before commercialisation requires very careful planning. There are precedents for Governments simply outsourcing everything, and it leads to poor management \cite{jenkin11}. Managing the process to commercialise fusion energy will require an approach that is more than that of just an `intelligent business' that knows what it needs. Experience suggests that an `intelligent customer' will be needed that includes an extremely capable in-house capability, that can  manage procurement, understand opportunities and potential innovation on the horizon, as well as integrate the programme into (changing) overall policies and structures \cite{chassels11}. 
Of course the scientific challenges addressed in this paper and required to develop commercial fusion are huge. But perhaps as challenging is developing an in-house management environment to roll-out fusion that attracts the required calibre of personnel with the relevant scientific, financial and administrative skills, and that provides a clear career development path for early career staff while retaining and developing its in-house expertise over a protracted capital-rich period of investment \cite{jenkin11}.

\begin{figure}
    \centerline{
    \includegraphics[height=15cm,keepaspectratio]{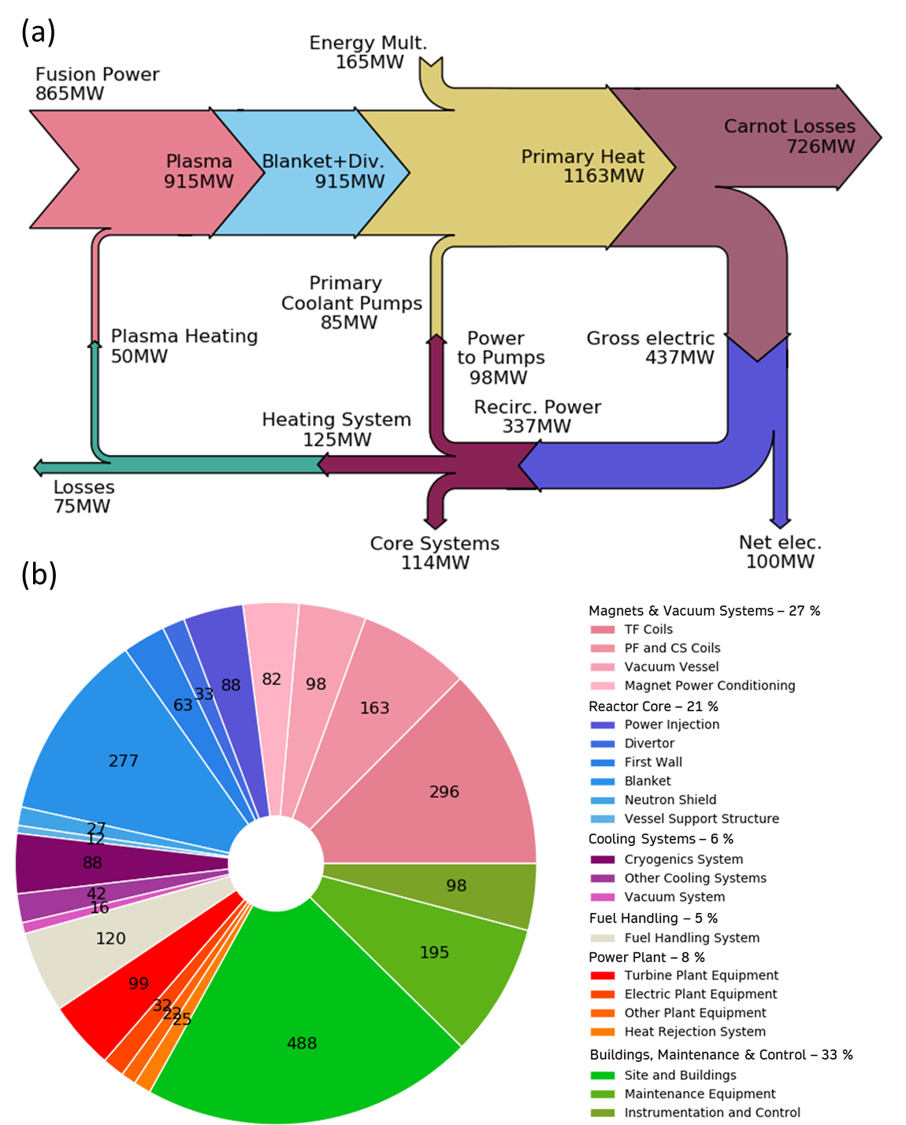}}
    \caption{\label{Power&Cost}(a) Reactor power balance (where ``Core Systems'' includes the cryo-system (46 \si{\mega\watt}$_\text{e}$) and the tritium handling system (15 \si{\mega\watt}$_\text{e}$)); (b) direct capital cost breakdown of our preferred 100 \si{\mega\watt} net electricity producing REBCO based tokamak power plant. Costs are in 1990 US M\$.}
\end{figure}

In this paper we focus on our `preferred choice' 100 \si{\mega\watt}$_\text{e}$ REBCO TF,CS and Nb-Ti PF tokamak with a plasma fusion gain $Q_\text{plasma}$ = 17, a net gain $Q_\text{net}$ = 1.3. The machine it is most similar to in size is ITER (cf Table \ref{ReactorComparison}). In a competitive commercial environment, the best machine must be sufficiently close to market to de-risk all the key technologies, provide a working demonstrator and a cost for producing fusion energy that the markets can rely on. Analysis of the power balance and direct capital cost for this preferred reactor are shown in figure \ref{Power&Cost}. The magnets are a single-point of failure for the entire project so we have mitigated damage to the brittle and expensive magnets by recommending the use of Nb-Ti training coils for the preferred choice reactor. Such training coils would increase overall preferred reactor capital cost by $<$ 10\%, and would allow for the thorough testing of the reactor (at 70 \% field on TF coil 66 \% field on CS coil) - reducing the risk of damage to the full-power REBCO magnets. The training coils are re-usable, so the increase in cost can shared across say 10 tokamaks meaning the increased cost per tokamak is rather marginal at only 1 \% \cite{Cardozo16}. These considerations hold for machines both with higher H$_{98}$ = 1.6 and for the 500 \si{\mega\watt}$_\text{e}$ scale. The cost of electricity is expensive from our preferred choice 550 \$ \si{\per\mega\watt\per\hour}, compared to 50-100 \$ \si{\per\mega\watt\per\hour} for fossil fuels or solar/wind (in 1990 US \$) - equivalent to 1148 \$ \si{\per\mega\watt\per\hour} and 100-200 \$ \si{\per\mega\watt\per\hour} respectively in 2021 US\$ costs \cite{IRENA21}. However, the primary aim of this paper is to minimise the capital cost of de-risking commercial fusion  technology, not produce cheap electricity. The 500 \si{\mega\watt}$_\text{e}$ machine produces electricity at a lower cost of 290 \$ per \si{\mega\watt\hour}, a plasma fusion gain $Q_\text{plasma}$ = 41, a net gain $Q_\text{net}$ = 1.9, demonstrating the benefits of large scale plants \cite{lee15}.\par

We have identified two areas of research that should be prioritised by the fusion community: Measurements and theoretical calculations are required to evaluate the properties of superconductors shielded but exposed to a fusion-like flux as a matter of urgency. MCNP calculations have been extended down to photon energies of 1 eV \cite{goorley16}, but need to be extended to consider even lower energies of order the superconducting gap (of order 10 meV). Perhaps even more challenging will then be developing the calculations and understanding of the superconducting state in such a photon and neutron flux; The development of high $B_{c2}$ multifilamentary superconducting alloys that will bring ductility, more straightforward robotic handling and maintenance of the magnets, and enhanced radiation tolerance (without requiring very high critical current density), opens the possibility to eliminate the need in large commercial machines to use the brittle HTS superconducting materials. We have concluded that superconducting magnets are the enabling technology for commercialisation of fusion power and that resistive magnets on any large scale are an unhelpful diversion of resources.

Finally we return to consider the global imperatives of net zero carbon economies with huge base-load electricity needs \cite{IEA2021}. These requirements have an obvious solution using nuclear power. Nuclear fission plants are currently the only commercial nuclear option \cite{Corkhill18}, but bring with them a current estimated cost of £132 billion (UK 2020) to decommission the UK’s current civil nuclear waste \cite{UK_decomissioning}. We haven't costed managing the nuclear waste for fusion power here, but support the intention for fusion technology to have no high level waste after 100 years (i.e. after the iron and cobalt isotopes have decayed) \cite{gilbert17,gilbert18,gilbert19,Bailey21} which brings public support for nuclear fusion power with it and remains essential \cite{turcanu20}.

\section{Conclusions}\label{Conclusions}
 
Prototype fusion reactors must demonstrate a commercial viability that proves fusion is a key option for large scale decarbonisation of global electricity production. To this end, here we have used the \texttt{PROCESS} systems code to minimise the capital cost of superconducting tokamak pilot plants using best-in-class technologies available on the time scales of 2035-2040. Should the cost of REBCO continue to fall and reach the expected 30 \$/\si{\kilo\ampere\meter} (6 \si{\tesla}, 4.2 K) \cite{cooley16}, our preferred reactor which uses REBCO CS and TF coils and Nb-Ti PF coils has a lower capital cost than a reactor with Nb$_3$Sn CS and TF coils, or an entirely Nb-Ti reactor. This preferred reactor has $R_\text{major}$ = 6.75 \si{\meter}, $A$ = 3.15, $P_\text{fusion}$ = 870 \si{\mega\watt}, $Q_{plasma}$ = 17, $Q_{net}$ = 1.3, $B_T$ = 5.4 \si{\tesla} and $B_\text{t,coil}^\text{max}$ = 12.5 \si{\tesla}. The cost of the preferred REBCO power plant is \$ 4230 M (US 1990) (cf Table 7), but if we use training magnets the total cost only increases to \$ 4.54 Bn (US 1990) equivalent to \$ 9.75 Bn (US 2021). We find the cost-optimal winding pack current density and field on coil is limited by the yield stress of the steel support structure and not the critical current density of the REBCO superconductor.
In order to achieve the necessarily high availability of a base-load power plant we assert that commercial tokamaks must have remountable magnets, and hence a prototype reactor must also demonstrate them. Remountability offers the possibility of easily swapping superconductors at different stages during a reactor lifetime. Here we have focused on using robust Nb-Ti training coils during plant commissioning in order to protect the brittle REBCO magnets from powerful disruptions as a result of operator inexperience, manufacturing errors etc. The full power magnets would then be mounted for full-power operation. The addition of Nb-Ti training coils for reactor commissioning increases costs by $<$ 10 \% and would allow for thorough reactor testing (generating 70 \% field on TF, 66 \% field on CS of the preferred reactor). We also suggest that the fusion community should commit itself to: developing higher $B_\text{c2}$ alloys dedicated to fusion applications \cite{horiuchi73,collings86}. They would aim to displace current commercial Nb-Ti, and because of their ductility also may displace high temperature superconductors  for fusion magnets; measure the critical current density of superconductors under operational conditions (bathed in an n+p sea at cryogenic temperatures).  

We conclude: the cost of building the human resources, engineering processes, supply chains, and capital-intensive demonstrator power plant that is the last one before commercialisation will not change significantly (against inflation) over the next few decades; all the relevant information is available now for the complex scientific and economic choices to be made about de-risking and then building a state-of-the-art tokamak. It should include all the key technologies and identify whether fission power plants can be replaced on a commercial footing by fusion power plants that produce no long-term high-level waste, reduce proliferation of nuclear weapons in a increasingly carbon-free world, and provide long-term energy security for base load electricity production.

\section{Acknowledgements} \label{Acknowledgements}
This work is funded by EPSRC grant EP/L01663X/1 that supports the EPSRC Centre for Doctoral Training in the Science and Technology of Fusion Energy. The data are available at: http://dx.doi.org/10.15128/r24q77fr362 and associated materials are on the Durham Research Online website:  http://dro.dur.ac.uk/.  We would like to thank, A. Turner and J. Naish for the MCNP calculations, as well as M. Kovari, J. Morris, S. Muldrew, S. Kahn, A. Blair, M. Coleman, P. Daniels, P. Bruzzone, K. Sedlak, S. Thomas, B. Davies, S. Frank, F. Schoofs, J. Schwartz, and also M. Raine, J. Greenwood, C. Gurnham and B. Din for their expertise and helpful discussions.

\pagebreak 

\bigskip
\section*{References} \label{References}
\bibliography{References.bib}

\clearpage

\begin{landscape}

\begin{table}
\caption{\label{ReactorComparison} Key design and performance parameters of \texttt{PROCESS} generated 100 \si{\mega\watt} net electricity (\si{\mega\watt}$_\text{e}$) and 500 \si{\mega\watt} net electricity (\si{\mega\watt}$_\text{e}$) tokamaks. Our preferred REBCO tokamak is shown in {\bf{bold}}. Also shown are: EU-DEMO \cite{federici18}, ARIES-ST \cite{najmabadi03}, a \texttt{PROCESS} generated pulsed Cu reactor based on STPP \cite{wilson04}, ARC \cite{sorbom15}, CFETR \cite{zhuang19,liu20}, ITER \cite{aymar02,sborchia11}, SPARC \cite{creely20,greenwald18}, JET \cite{bertolini00,romanelli15,pamela03}, JT60-SA \cite{JT60mags,shirai17}, KSTAR \cite{lee01,kim05c,ahn15}, EAST \cite{shen09,weng05,liu12}, WEST \cite{bourdelle15,duchateau07}, MAST-U \cite{menard16,harrison19} and SST-1 \cite{bora02,saxena04}. Tokamaks have been grouped into: those in this work, demonstration reactors, proof of concept reactors and research tokamaks. Estimated parameters indicated with (*). Tokamaks with resistive primary magnets are indicated with ($\dagger$).
}

\begin{indented}\lineup
    \lineup
    \item[]\begin{tabular}{@{}clllllllllll} 
    \br
    &\makecell[l]{Tokamak\\\0}&\makecell[l]{H$_\text{98}$\\\0}&\makecell[l]{$R_\text{major}$\\(m)}&\makecell[l]{$R_\text{minor}$\\(m)}&\makecell[l]{$B_\text{plasma}$\\(T)}&\makecell[l]{$B_\text{coil}^\text{max}$\\(T)}&\makecell[l]{$I_\text{P}$\\(MA)}&\makecell[l]{$\tau_\text{burn}$\\(s)}&\makecell[l]{$P_\text{fusion}$\\(\si{\mega\watt})}&\makecell[l]{$P_\text{elec.}^\text{net}$\\(\si{\mega\watt})}\\
    \br
    \parbox[t]{2mm}{\multirow{6}{*}{\rotatebox[origin=c]{90}{In this work}}}&\makecell[l]{\bf{100 MW$_\text{e}$ REBCO plant}}& \bf{1.2} &\bf{6.75} &\bf{2.14} &\0\bf{5.36} &\bf{12.50}  &\bf{13.6}   &\bf{7200}  &\0\bf{870}    &\0\bf{100}\\
    &\makecell[l]{100 \si{\mega\watt}$_\text{e}$ REBCO plant}& 1.6 & 6.02& 1.85 &\05.22 & 12.63 &10.9  &7200 &\0840   &\0100\\
    &\makecell[l]{500 \si{\mega\watt}$_\text{e}$ REBCO plant}& 1.2 &7.48 &3.13 &\04.18 &11.85  &25.2   &7200   &2110    &\0500\\
    &\makecell[l]{100 \si{\mega\watt}$_\text{e}$ Nb-Ti plant}& 1.2 &7.93 &3.05 &\03.38 &\09.16 &18.9   &7200  &\0900    &\0100\\
    &\makecell[l]{100 \si{\mega\watt}$_\text{e}$ Nb-Ti plant}& 1.6 &6.96&2.95 &\02.96&\09.00&15.9  &7200 &\0870   &\0100\\
    &\makecell[l]{500 \si{\mega\watt}$_\text{e}$ Nb-Ti plant}& 1.2 &9.56 &4.27 &\03.41 &\09.17 &29.4   &7200   &2180    &\0500\\

    \mr
    \parbox[t]{2mm}{\multirow{3}{*}{\rotatebox[origin=c]{90}{Demo.}}}
    &\makecell[l]{ARC}&               1.8 &3.30 &1.13 &\09.20 &23.00  &\07.8  &\0\0$\infty$     &\0525     &\0190\\
    &\makecell[l]{CFETR}&            1.4&7.20&2.20 &\06.50 &    14.70 &13.8 &\0\0$\infty$&2190&740\\
    &\makecell[l]{EU-DEMO}&           1.1 &9.00 &2.90 &\05.90 &12.50  &18.0   &7200   &2000    &\0500\\
    \mr
    \parbox[t]{2mm}{\multirow{4}{*}{\rotatebox[origin=c]{90}{P.o.C}}}
    &\makecell[l]{SPARC}&             1.0 &3.25 &0.57 &12.20  &    20.00 &\08.7  &\0\010 &\0140    
    &\0\0-\\
    &\makecell[l]{ITER}&              1.0 &6.20 &2.00 &\05.30 &11.80  &15.0   &\0400  &\0500   &\0\0-\\
    &\makecell[l]{STPP-Like$\dagger$}&             1.6 &3.42 &2.04 &\02.50 &\07.56 &19.4   &7200    &2110    &\0100\\
    &\makecell[l]{ARIES-ST$\dagger$}&             1.5 &3.20 &2.00 &\02.10 &\07.40 &29.0   &\0\0$\infty$     &2980    &1000\\
    \mr
    \parbox[t]{2mm}{\multirow{7}{*}{\rotatebox[origin=c]{90}{Research}}}
     &\makecell[l]{WEST}&            1.0&2.50&0.50&\03.70&    \09.00 &\00.6 &1000&\0\0-&\0\0-\\
    &\makecell[l]{KSTAR}&             0.7-1.0 &1.80 &0.50 &\03.50 &    \07.20&\02.0  &\0\020 &\0\0-       &\0\0- \\
    &\makecell[l]{EAST}&            0.5-1.2&1.75&0.40 &\03.50 &  \05.80 &\01.0 &1000&\0\0-&\0\0-\\
   &\makecell[l]{JET$\dagger$}&               0.5-1.3 &2.96 &1.23 &\03.45 &  \07.40 &\04.8  &\0\0\01&\0\016  &\0\0-\\
   &\makecell[l]{SST-1}&             1.0-2.0 &1.10 &0.20 &\03.00 &    \05.10 &\00.2  &1000     &\0\0-       & \0\0- \\
   &\makecell[l]{JT60-SA}&       1.1-1.3 &3.40 &1.36 &\02.25 &\06.40 &\05.5 &\0100  &\0\041  &\0\0-\\ &\makecell[l]{MAST-U$\dagger$}&            1.0-2.0&0.85&0.65 &\00.92 &    \04.20 &\01.0 &\0\0\05&\0\0-&\0\0-\\

     \br
    \end{tabular}
\end{indented}
\end{table}
\end{landscape}

\begin{table}[h]
\caption{\label{Cost_Breakdown} Capital cost of the preferred reactor and two other \texttt{PROCESS} generated, cost-optimised, 100 \si{\mega\watt} net electricity, H$_\text{98}$ = 1.2 tokamak pilot plants. Bold - preferred reactor with REBCO TF and CS coils. Also shown are plants with: Nb$_3$Sn TF and CS coils; Commercial Nb-Ti TF and CS coils and quaternary Nb-Ti TF and CS coils. In all cases the PF coils are Nb-Ti. Note that these costs are for simply building the plant without mitigating risk with training or upgrading coils. Plant direct cost is the cost of the buildings, raw materials and labour only. Plant constructed costs also include the indirect costs (R\&D, admin, licensing etc.), contingency costs and budget loan repayments. Total capital investment is the plant constructed cost as well as accrued interest on loan repayments and loss of buying power from the budget due to inflation during construction. All costs are in 1990 M\$.}

\begin{indented}\lineup
\item[]\begin{tabular}{@{}lllll}
\toprule
& \makecell[l]{\bf{Preferred}\\\bf{(REBCO)}\\\bf{reactor}} &\makecell[l]{Nb$_3$Sn\\plant}& \makecell[l]{Commercial\\Nb-Ti\\plant}&\makecell[l]{Quaternary\\Nb-Ti\\plant}\\
\midrule
Structures and site facilities     &\0\bf{489}&\0530&\0667&\0546\\
Reactor systems	                   &\0\0\bf{39}&\0\045&\0\061&\0\047\\
Toroidal field coils               &\0\bf{296}&\0315&\0243&\0296\\
\makecell[r]{(TF cable)}           &\bf{(130)}&\0(98)&\0(62)&\0(75)\\
Poloidal field coils and solenoid  &\0\bf{163}&\0159&\0181&\0153\\
\makecell[r]{(CS cable)}           &\0\bf{(20)}&\0(19)&\0(20)&\0(17)\\
\makecell[r]{(PF cable)}           &\0\bf{(79)}&\0(74)&\0(80)&\0(69)\\ 
First wall                         &\0\0\bf{63}&\0\073&\0107&\0\077\\
Blanket                            &\0\bf{277}&\0321&\0479&\0341\\
Divertor                           &\0\0\bf{33}&\0\038&\0\059&\0\041\\
Heating \& current drive           &\0\0\bf{10}&\0\010&\0\010&\0\010\\
Vacuum vessel                      &\0\0\bf{98}&\0112&\0157&\0118\\
Power injection	                   &\0\0\bf{88}&\0\088&\0\088&\0\088\\
Vacuum systems 	                   &\0\0\bf{16}&\0\016&\0\016&\0\016\\
Power conditioning	               &\0\0\bf{82}&\0\088&\0\089&\0\077\\
Heat transport system 	           &\0\bf{130}&\0139&\0130&\0127\\
\makecell[r]{(Cryogenics system)}  &\0\bf{(88)}&\0(95)&\0(86)&\0(85)\\
Fuel handling system	           &\0\bf{120}&\0127&\0147&\0129\\
Instrumentation and control        &\0\0\bf{98}&\0\098&\0\098&\0\098\\
Maintenance equipment	           &\0\bf{195}&\0195&\0195&\0195\\
Turbine plant Equipment            &\0\0\bf{99}&\0102&\0103&\0\099\\
Electric plant equipment           &\0\0\bf{32}&\0\034&\0\038&\0\034\\
Miscellaneous plant equipment	   &\0\0\bf{22}&\0\022&\0\022&\0\022\\
Heat rejection system              &\0\0\bf{25}&\0\026&\0\026&\0\025\\
Plant direct cost                  &\bf{2373}&2536&2915&2539\\
Constructed cost                   &\bf{3631}&3881&4462&3885\\
\midrule
Total Capital Investment (1990 M\$)&\bf{4231}&4522&5198&4526\\
\bottomrule
\end{tabular}
\end{indented}
\end{table}

\begin{landscape}
\begin{table}
\caption{\label{Power_Breakdown}Power balance of the preferred reactor and two other \texttt{PROCESS} generated, cost-optimised, 100 \si{\mega\watt} net electricity, H$_\text{98}$ = 1.2 tokamak pilot plants. Bold - preferred reactor with REBCO TF and CS coils. Also shown are plants with: Nb$_3$Sn TF and CS coils; Commercial Nb-Ti TF and CS coils and quaternary Nb-Ti TF and CS coils. In all cases the PF coils are Nb-Ti.}
\begin{indented}\lineup
\item[]\begin{tabular}{@{}llllll}
\toprule
    && \makecell[l]{\bf{Preferred}\\\bf{(REBCO)}\\\bf{reactor}} &\makecell[l]{Nb$_3$Sn\\plant}& \makecell[l]{Commercial\\Nb-Ti\\plant}&\makecell[l]{Quaternary\\Nb-Ti\\plant}\\
        \toprule
Raw Heat       & Fusion power                &\bf{865}&889&902&860\\
(M\si{\watt})  & Blanket multiplication      &\bf{165}&169&172&164\\
               & Injected power              &\bf{50}&50&50&50\\
               & Ohmic heating               &\bf{1}&1&1&1\\
               & Power from coolant pump     &\bf{85}&87&88&84\\ \cline{3-6} 
               &                             &\bf{1166}&1196&1213&1158\\
\midrule
Gross electric & Power lost in conversion    &\bf{-729}&-747&-758&-724\\ \cline{3-6}  
power (M\si{\watt}) &                        &\bf{437}&449&455&434\\ 
\midrule
Net electric   & Heating and current drive   &\bf{-125}&-125&-125&-125\\
power (M\si{\watt}) & Primary coolant pumps  &\bf{-98}&-100&-101&-97\\
               & Vacuum pumps                &\bf{-1}&-1&-1&-1\\
               & Tritium plant               &\bf{-15}&-15&-15&-15\\
               & Cryoplant                   &\bf{-44}&-50&-43&-42\\
               & Toroidal field coils        &\bf{-12}&-12&-13&-8\\
               & Poloidal field coils and solenoid &\bf{-1}&-1&-1&-1\\
               & Miscellaneous               &\bf{-42}&-45&-57&-47\\ \cline{3-6}  
               &                             &\bf{100}&100&100&100\\
\bottomrule
\end{tabular}
\end{indented}
\end{table}
\end{landscape}

\begin{table}
\caption{\label{mus} Mean attenuation coefficients for (fast) neutrons of energy $>$ 0.1 \si{\mega\electronvolt} of tokamak relevant materials. $\overline{\mu}_\text{TCA}$ are calculated from total neutron cross section data \cite{ENDF}. $\overline{\mu}_\text{i}$ calculated using data from \texttt{MCNP} calculations of neutron transmission through a 30 \si{\centi\meter} block of (ith) mono-material \cite{colling16} except for Tungsten Carbide which is derived from the \texttt{MCNP} data in figure \ref{Xplot}. }
\begin{indented}
\lineup
\item[]\begin{tabular}{@{}lll} 
\br
Material&$\makecell[l]{\overline{\mu}_\text{TCA}(E > 0.1 \si{\mega\electronvolt})\\(\si{\per\meter})}$&$\makecell[l]{\overline{\mu}_\text{i}(E > 0.1 \si{\mega\electronvolt})\\(\si{\per\meter})}$\\
\br
Tungsten	            &42.71	&19.55\\
304B7 Boronated Steel 	&39.92	&16.44\\
316 Stainless Steel     &40.06  &15.31\\
Copper	                &34.71	&14.98\\   
Niobium	                &34.77	&13.73\\
Beryllium	            &39.97	&14.61\\
Tin	                    &15.75	&12.06\\
Zirconium	            &26.26	&11.87\\
Gadolinium	            &19.85	&12.63\\
Titanium	            &15.75	&13.01\\
Water	                &13.73	&\08.15\\
Aluminium               &21.43	&13.53\\
Lithium	                &13.16	&10.72\\
Helium (liquid)	        &\04.56	&\09.45\\
Hydrogen                &\01.94	&\02.45\\
Tungsten Carbide        &23.06	&\018.9 \\

\br
\end{tabular}
\end{indented}
\end{table}

\begin{table}
\caption{\label{ShieldThicknesses}Thicknesses and material compositions (derived from the ITER radial build) of the layers between the first wall and the central solenoid at the inboard mid-plane for the initial 100 \si{\mega\watt} net electricity REBCO CS, TF and Nb-Ti PF reactor using a radiation shield thickness from the benchmarking calculations. As well as those for the preferred reactor (in bold) using a radiation shield thickness optimised using \texttt{MCNP}. Outboard dimensions are shown in brackets () where significantly different.}
\begin{indented}
\lineup
\item[]\begin{tabular}{@{}lllll} 
\br
Section & Layer & \makecell[l]{Material\\ composition}& \makecell[l]{initial component\\ thicknesses\\ (m)}& \makecell[l]{\bf{preferred design}\\ \bf{thicknesses}\\ \bf{(m)}}\\\ns
\br
First wall & 	Armour& Tungsten&	0.010&	\bf{0.010} \\\cline{2-3}
&Cooling& 	\makecell[l]{90\% Glidcop,\\ 10\% water}&	0.008 &	\bf{0.008} \\
\mr
\makecell[l]{Breeder\\blanket} && \makecell[l]{	37.5\% TiBe$_\text{12}$\\ 37.5\% Li$_2$SiO$_4$\\9.7\% 316 stainless steel\\ 15.3 \% He}&	0.530 (0.910)	& \bf{0.530 (0.910)}\\
\mr
Gap&&	 	Air&	0.010 &	\bf{0.010} \\
\mr
Radiation shield	 &	&Tungsten Carbide&	0.214 &	\bf{0.250} \\
\mr
Vacuum vessel &	Wall& 	316 stainless steel&	0.060 &	\bf{0.060} \\\cline{2-3}
&Interior&	\makecell[l]{60\% 304 stainless steel\\ with 2\% Boron\\ 40\% water}&	0.200 (0.350)&	\bf{0.200 (0.350)}\\\cline{2-3}
&Wall &	316 stainless steel&	0.060 &	\bf{0.060} \\
\mr
Gap&&	 	Air& 	0.010 &	\bf{0.010} \\
\mr
Thermal shield	&& 	316 stainless steel&	0.050 &	\bf{0.050} \\
\mr
Gap	 	&&Air&	0.065 (0.75)&	\bf{0.065 (0.75)}\\
\mr
TF coil &	TF coil casing&316 stainless steel&0.050 &	\bf{0.052} \\\cline{2-3}
 	&Insulation& \makecell[l]{45\% Fibreglass tape\\ 45\% Kapton tape\\ 10\% epoxy resin}&	0.018 &	\bf{0.018} \\\cline{2-3}
 	&Winding pack& \makecell[l]{$<$1\% REBCO\\ 51\% Copper\\ 28\% Hastelloy\\ 20\% He}&	0.460 &	\bf{0.550} \\\cline{2-3}
 &	Insulation	&\makecell[l]{45\% Fibreglass tape\\ 45\% Kapton tape\\ 10\% epoxy resin}&	0.018 &	\bf{0.018} \\\cline{2-3}
 &	TF coil casing&316 stainless steel&	0.070 &	\bf{0.070} \\
\br

\end{tabular}
\end{indented}
\end{table}

%\begin{landscape}
\begin{table}
\caption{\label{Cryoproperties} Above: Useful cryogenic materials properties under 5 \si{\bar} pressure \cite{TPFS}. Below: (magneto)resistances of RRR = 1000 copper and RRR = 10000 aluminium \cite{simon92,egan90,fickett72}. }
\begin{indented}\lineup
\item[]\begin{tabular}{@{}lllll} 
    \br
&\makecell[l]{$c_p$ \\ (\si{\joule\per\kilogram\per\kelvin})}&\makecell[l]{$\kappa$ \\ (\si{\watt\per\meter\per\kelvin})}&\makecell[l]{$\rho$\\ (\si{\kilogram\per\cubic\meter})}&\makecell[l]{$\mu$\\(\si{\mega\pascal\second})}\\
\mr
Water (293 \si{\kelvin})&4183&0.598&\0998&1005\\
N$_2$  (65 \si{\kelvin})&1165&0.174&\0861&\0282\\
Ne (30 \si{\kelvin})&2009&0.138&1156&\0\089\\
He (30 \si{\kelvin})&5312&0.034&\0\0\08&\0\0\05\\
He (20 \si{\kelvin})&5472&0.027&\0\012&\0\0\04\\
He (4.5 \si{\kelvin})&3955&0.021&\0143&\0\0\04\\
    \br
    \end{tabular}
\end{indented}

\begin{indented}
\item[]\begin{tabular}{@{}lll}
\br
&\makecell[l]{Cu Resistivity\\ (\si{\ohm\meter})}&\makecell[l]{Al Resistivity\\ (\si{\ohm\meter})}\\
\mr
293 \si{\kelvin}, 0 \si{\tesla}&1.7$\times10^{-8}$&2.7$\times10^{-8}$\\
65 \si{\kelvin}, 0 \si{\tesla}&1.5$\times10^{-9}$&1.1$\times10^{-9}$\\
65 \si{\kelvin}, 6 \si{\tesla}&9.9$\times10^{-9}$&2.9$\times10^{-9}$\\
20 \si{\kelvin}, 0 \si{\tesla}&2.8$\times10^{-11}$&9.0$\times10^{-12}$\\
20 \si{\kelvin}, 6 \si{\tesla}&3.1$\times10^{-10}$&6.8$\times10^{-11}$\\
\br
\end{tabular}
\end{indented}

\end{table}
%\end{landscape}

\begin{table}
\caption{\label{JcScaling}Parameters for the Durham scaling law for transport critical current measurements on ITER specification Nb-Ti strands, quaternary Nb-Ti (not available commercially in long lengths), internal-tin Nb$_3$Sn and REBCO. Extensive measurements have yielded that $B_\text{c2}(T) = B_\text{c2}(0)(1-t^\nu),$ for low temperature superconductors \cite{chislett-mcdonald20a,taylor05,keys03,lu08}, and $B_\text{c2}(T) = B_\text{c2}(0)(1-t)^s~$ for REBCO \cite{branch20}. The ITER spec. Nb-Ti parameters were found by extensive measurements taken at Durham university. $B_\text{c2}^*(0)$ and $T_\text{c}^*$ for quaternary Nb-Ti were taken from \cite{horiuchi73}, all other parameters are from ITER specification Nb-Ti. The value of $A^*$ for REBCO was found by fitting to literature data \cite{braccini10}, all other parameters were taken from measurements on REBCO tapes \cite{branch20}. Nb$_3$Sn parameters were taken from \cite{taylor05}.}
\begin{indented}\lineup
\item[]\begin{tabular}{@{}lllllllll} 
\br
&\makecell[l]{$A^*$\\(\si{\ampere\meter^{-2}  \kelvin^{-2} \tesla^{3-\it{n}}})}& \makecell[l]{$T_\text{c}^*$\\(\si{\kelvin})} & \makecell[l]{$B_\text{c2}^*(0)$\\(\si{\tesla})}&$p$&$q$&$n$&$v$&$s$\\
\br
\makecell[l]{Comm.\\ Nb-Ti}&3.42$\times10^8$&\0\09.04&\014.86&0.49&0.56&1.83&1.42&~~-\\
\makecell[l]{Quat.\\Nb-Ti}&3.42$\times10^8$&\0\08.30&\021.13&0.49&0.56&1.83&1.42&~~-\\
\makecell[l]{REBCO\\}&1.24$\times10^3$&184.98&138.97&0.45&1.44&3.33&~~-&5.27\\
\makecell[l]{Nb$_3$Sn\\}&2.45$\times10^7$&\016.89&\028.54&0.47&1.95&2.34&1.45&~~-\\
\br
\end{tabular}
\end{indented}
\begin{indented}
\item[]\begin{tabular}{lllllllll} 
\br
&$v$&$s$&$u$&$w$&$c_2$&$c_3$&$c_4$&\makecell[l]{$\varepsilon_{m}$\\(\%)}\\
\mr
\makecell[l]{Comm.\\Nb-Ti}&1.42&~~-&0.00&2.2&-0.0025&-0.0003&-0.0001&-0.002\\
\makecell[l]{Quat.\\Nb-Ti}&1.42&~~-&0.00&2.2&-0.0025&-0.0003&-0.0001&-0.002\\
\makecell[l]{REBCO\\}&~~-&5.27&0.00&2.2&-0.0191&0.0039&0.00103&0.058\\ 
\makecell[l]{Nb$_3$Sn\\}&1.45&~~-&-0.06&1.94&-0.7697&-0.4913&-0.0538&0.279\\
\br
\end{tabular}
\end{indented}
\end{table}
\clearpage

\begin{landscape}
\begin{table}
\caption{\label{TrainingH=1.2}Trained tokamaks designed for 100 \si{\mega\watt} net electricity and H$_\text{98}$ = 1.2: Performance and cost data for the (preferred) tokamak optimised for REBCO toroidal field and central solenoid coils with minimised capital cost and training magnets of different superconductors (with maximised net electricity yield). The different superconductors considered are the high temperature superconductor $REBa_2Cu_3O_{7}$ (REBCO where $RE$:rare-earth), commercial NbTi (Comm. NbTi) used in MRI scanners, and quaternary NbTi (Quat. NbTi) that is not yet available commercially. In bold are the preferred tokamak (\$ 4230 M) and the preferred tokamak with commercial Nb-Ti training magnets (\$ 4540 M). Also shown are the power values that would result, were a reduced H$_\text{98}$ = 1.0 to occur in practice.}
\begin{indented}\lineup
\item[]\begin{tabular}{@{}llllllllllll} 
    \br
    &&&&&&&& \multicolumn{2}{c}{H$_\text{98}$ = 1.2} & \multicolumn{2}{c}{H$_\text{98}$ = 1.0}\\\cline{9-12}
    
    TF s.c.& CS s.c.&\makecell[l]{TF (CS)\\ peak field\\(\si{\tesla})}&\makecell[l]{Field on\\plasma\\(\si{\tesla})}&\makecell[l]{TF (CS)\\steel \%\\(\%)}&\makecell[l]{TF (CS)\\peak stress\\ (\si{\mega\pascal})}&\makecell[l]{Plasma\\current\\(M\si{\ampere})}&\makecell[l]{Capital\\cost\\(1990 M\$)}&\makecell[l]{Fusion\\power\\(M\si{\watt})}&\makecell[l]{Net.\\elec.\\(M\si{\watt})}&\makecell[l]{Fusion\\power\\(M\si{\watt})}&\makecell[l]{Net.\\elec.\\(M\si{\watt})}\\
    \br
    										
   \bf{REBCO}&\bf{REBCO}&\bf{12.5 (14.0)}&\bf{5.4}&\bf{52.8 (80.4)}&\bf{660 (660)}&\bf{13.6}&\bf{4230}&\bf{860}&\bf{100}&\bf{20}&\bf{-190}\\
    \mr
    							
    \makecell[l]{Comm.\\Nb-Ti}&REBCO&8.7 (14.0)&3.7&38.5 (80.4)&390 (450)&9.5&4580&90&-170&30&-190\\
    										
    \makecell[l]{Quat.\\Nb-Ti}&REBCO&11.6 (14.0)&5.0&47.1 (80.4)&600 (660)&12.6&4630&530&-20&30&-190\\
    										
    \makecell[l]{\bf{Comm.}\\\bf{Nb-Ti}}&\makecell[l]{\bf{Comm.}\\\bf{Nb-Ti}}&\bf{8.8 (9.2)}&\bf{3.7}&\bf{44.9 (42.0)}&\bf{560 (620)}&\bf{8.6}&\bf{4540}&\bf{60}&\bf{-180}&\bf{30}&\bf{-190}\\
    										
    \makecell[l]{Quat.\\Nb-Ti}&\makecell[l]{Quat.\\Nb-Ti}&11.0 (9.9)&4.7&47.1 (69.4)&660 (620)&11.0&4620&130&-150&40&-170\\
    										
   \br
    \end{tabular}
\end{indented}
\end{table}
\end{landscape}

\begin{landscape}
\begin{table} 
\caption{\label{TrainingH=1.6}
Trained tokamaks designed for 100 \si{\mega\watt} net electricity and H$_\text{98}$ = 1.6: Performance and cost data for a tokamak optimised for REBCO toroidal field and central solenoid coils with minimised capital cost (*) and training magnets of different superconductors (with maximised net electricity yield). Also shown are the power values that would result, were a reduced H$_\text{98}$ = 1.0 to occur in practice.}
\begin{indented}\lineup
\item[]\begin{tabular}{@{}lllllllllllll} 
    \br
    &&&&&&&& \multicolumn{2}{c}{H$_\text{98}$ = 1.6} & \multicolumn{2}{c}{H$_\text{98}$ = 1.0}\\\cline{9-12}
    
    TF s.c.& CS s.c.&\makecell[l]{TF (CS)\\ peak field\\(\si{\tesla})}&\makecell[l]{Field on\\plasma\\(\si{\tesla})}&\makecell[l]{TF (CS)\\steel \%\\(\%)}&\makecell[l]{TF (CS)\\peak stress\\ (\si{\mega\pascal})}&\makecell[l]{Plasma\\current\\(M\si{\ampere})}&\makecell[l]{Capital\\cost\\(1990 M\$)}&\makecell[l]{Fusion\\power\\(M\si{\watt})}&\makecell[l]{Net.\\elec.\\(M\si{\watt})}&\makecell[l]{Fusion\\power\\(M\si{\watt})}&\makecell[l]{Net.\\elec.\\(M\si{\watt})}\\
    \br
    
    REBCO*&REBCO*&12.6 (11.9)&5.2&52.0 (81.2)&660 (660)&10.9&3610&830&100&30&-190\\
    \mr
    							
    \makecell[l]{Comm.\\Nb-Ti}&REBCO&9.1 (11.2)&3.7&41.3 (81.2)&410 (240)&7.4&3920&140&-150&20&-190\\
    										
    \makecell[l]{Quat.\\Nb-Ti}&REBCO&11.8 (11.6)&5.0&48.5 (81.2)&590 (550)&9.2&3950&420&-50&30&-190\\
    										
    \makecell[l]{Comm.\\Nb-Ti}&\makecell[l]{Comm.\\Nb-Ti}&9.3 (9.2)&4.0&52.3 (34.1)&450 (550)&7.3&4060&180&-130&30&-180\\
    										
    \makecell[l]{Quat.\\Nb-Ti}&\makecell[l]{Quat.\\Nb-Ti}&11.9 (10.3)&5.1&48.9 (73.2)&650 (580)&9.4&4010&560&0&20&-190\\
    											
    \br
\end{tabular}
\end{indented}
\end{table}
\end{landscape}

\begin{landscape}
\begin{table} 
\caption{\label{Training500}Trained tokamaks designed for 500 \si{\mega\watt} net electricity and H$_\text{98}$ = 1.2: Performance and cost data for a tokamak optimised for REBCO toroidal field and central solenoid coils with minimised capital cost (*) and training magnets of different superconductors (with maximised net electricity yield). Also shown are the power values that would result, were a reduced H$_\text{98}$ = 1.0 to occur in practice.}
\begin{indented}\lineup
\item[]\begin{tabular}{@{}lllllllllllll} 
    \br
    &&&&&&&& \multicolumn{2}{c}{H$_\text{98}$ = 1.2} & \multicolumn{2}{c}{H$_\text{98}$ = 1.0}\\\cline{9-12}
    
    TF s.c.& CS s.c.&\makecell[l]{TF (CS)\\ peak field\\(\si{\tesla})}&\makecell[l]{Field on\\plasma\\(\si{\tesla})}&\makecell[l]{TF (CS)\\steel \%\\(\%)}&\makecell[l]{TF (CS)\\peak stress\\ (\si{\mega\pascal})}&\makecell[l]{Plasma\\current\\(M\si{\ampere})}&\makecell[l]{Capital\\cost\\(1990 M\$)}&\makecell[l]{Fusion\\power\\(M\si{\watt})}&\makecell[l]{Net.\\elec.\\(M\si{\watt})}&\makecell[l]{Fusion\\power\\(M\si{\watt})}&\makecell[l]{Net.\\elec.\\(M\si{\watt})}\\
    \br
    
     REBCO*&REBCO*&11.9 (12.8)&4.2&51.9 (89.4)&660 (660)&25.3&5890&2120&500&1490&290\\
    \mr
    							
    \makecell[l]{Comm.\\Nb-Ti}&REBCO&9.1 (11.9)&3.2&52.7 (89.4)&260 (360)&179.5&6340&440&-70&220&-150\\
    										
    \makecell[l]{Quat.\\Nb-Ti}&REBCO&11.8 (12.6)&4.2&51.2 (89.4)&660 (660)&25.1&6460&2080&490&1490&290\\
    										
    \makecell[l]{Comm.\\Nb-Ti}&\makecell[l]{Comm.\\Nb-Ti}&9.3 (9.1)&3.3&43.2 (52.2)&590 (660)&19.8&6490&1060&150&620&-10\\
    										
    \makecell[l]{Quat.\\Nb-Ti}&\makecell[l]{Quat.\\Nb-Ti}&11.5 (11.0)&4.0&51.3 (72.8)&660 (660)&24.3&6500&1920&440&1250&210\\
    										
    \br
    \end{tabular}
\end{indented}
\end{table}
\end{landscape}

\begin{landscape}
\begin{table}
\caption{\label{UpgradeH=1.2}Upgraded tokamaks designed for 100 \si{\mega\watt} net electricity and H$_\text{98}$ = 1.2: Performance and cost data for a tokamak optimised for Nb-Ti toroidal field and central solenoid coils with minimised capital cost (*) and upgraded magnets of different superconductors (with maximised net electricity yield). Also shown are the power values that would result, were a reduced H$_\text{98}$ = 1.0 to occur in practice.}
\begin{indented}\lineup
\item[]\begin{tabular}{@{}lllllllllllll} 
    \br
    &&&&&&&& \multicolumn{2}{c}{H$_\text{98}$ = 1.2} & \multicolumn{2}{c}{H$_\text{98}$ = 1.0}\\\cline{9-12}
    
    TF s.c.& CS s.c.&\makecell[l]{TF (CS)\\ peak field\\(\si{\tesla})}&\makecell[l]{Field on\\plasma\\(\si{\tesla})}&\makecell[l]{TF (CS)\\steel \%\\(\%)}&\makecell[l]{TF (CS)\\peak stress\\ (\si{\mega\pascal})}&\makecell[l]{Plasma\\current\\(M\si{\ampere})}&\makecell[l]{Capital\\cost\\(1990 M\$)}&\makecell[l]{Fusion\\power\\(M\si{\watt})}&\makecell[l]{Net.\\elec.\\(M\si{\watt})}&\makecell[l]{Fusion\\power\\(M\si{\watt})}&\makecell[l]{Net.\\elec.\\(M\si{\watt})}\\
    \br
    
    \makecell[l]{Commer.*\\Nb-Ti}&\makecell[l]{Commer.*\\Nb-Ti}&9.2 (8.1)&3.4&45.2 (63.9)&660 (660)&18.9&5200&900&100&60&-200\\
    \mr
						
    \makecell[l]{Commer.\\Nb-Ti}&REBCO&9.2 (9.1)&3.4&45.2 (79.0)&660 (660)&19.3&5280&900&100&60&-200\\
    
    REBCO&REBCO&10.1 (13.2)&3.7&47.5 (91.5)&660 (660)&20.8&6470&1410&280&920&100\\
    										
    \makecell[l]{Quat.\\Nb-Ti}&\makecell[l]{Quat.\\Nb-Ti}&9.7 (10.3)&3.6&46.8 (79.2)&660 (660)&20.0&6150&1250&230&630&-40\\
    										
    \makecell[l]{Quat.\\Nb-Ti}&REBCO&10.1 (13.0)&3.7&47.5 (91.6)&660 (660)&20.8&6350&1400&280&900&100\\
    										
    \br
    \end{tabular}
\end{indented}
\end{table}
\end{landscape}

\begin{landscape}
\begin{table}
\caption{\label{UpgradeH=1.6}
Upgraded tokamaks designed for 100 \si{\mega\watt} net electricity and H$_\text{98}$ = 1.6: Performance and cost data for a tokamak optimised for Nb-Ti toroidal field and central solenoid coils with minimised capital cost (*) and upgraded magnets of different superconductors (with maximised net electricity yield). Also shown are the power values that would result, were a reduced H$_\text{98}$ = 1.0 to occur in practice.}
\begin{indented}\lineup
\item[]\begin{tabular}{@{}lllllllllllll} 
    \br
    &&&&&&&& \multicolumn{2}{c}{H$_\text{98}$ = 1.6} & \multicolumn{2}{c}{H$_\text{98}$ = 1.0}\\\cline{9-12}
    
    TF s.c.& CS s.c.&\makecell[l]{TF (CS)\\ peak field\\(\si{\tesla})}&\makecell[l]{Field on\\plasma\\(\si{\tesla})}&\makecell[l]{TF (CS)\\steel \%\\(\%)}&\makecell[l]{TF (CS)\\peak stress\\ (\si{\mega\pascal})}&\makecell[l]{Plasma\\current\\(M\si{\ampere})}&\makecell[l]{Capital\\cost\\(1990 M\$)}&\makecell[l]{Fusion\\power\\(M\si{\watt})}&\makecell[l]{Net.\\elec.\\(M\si{\watt})}&\makecell[l]{Fusion\\power\\(M\si{\watt})}&\makecell[l]{Net.\\elec.\\(M\si{\watt})}\\
    \br
    \makecell[l]{Commer.*\\Nb-Ti}&\makecell[l]{Commer.*\\Nb-Ti}&9.0 (7.3)&3.0&45.5 (58.8)&660 (660)&15.9&4470&870&100&20&-200\\
    \mr
						
    \makecell[l]{Commer.\\Nb-Ti}&REBCO&9.0 (7.2)&3.0& 45.5 (69.5)&660 (660)&15.9&4530&870&100&20&-200\\
    									
    REBCO&REBCO&10.2 (14.4)&3.3&46.7 (90.8)&660 (660)&18.1&5280&1190&210&30&-200\\
    										
    \makecell[l]{Quat.\\Nb-Ti}&\makecell[l]{Quat.\\Nb-Ti}&9.7 (10.9)&3.2&45.8 (71.0)&660 (660)&17.3&5180&1070&170&30&-200\\
    										
    \makecell[l]{Quat.\\Nb-Ti}&REBCO&10.2 (14.2)&3.4&46.7 (90.7)&660 (660)&18.1&4890&1200&210&30&-200\\
    
    \br
    \end{tabular}
\end{indented}
\end{table}
\end{landscape}

\begin{landscape}
\begin{table}
\caption{\label{Upgrade500}
Upgraded tokamaks designed for 500 \si{\mega\watt} net electricity and H$_\text{98}$ = 1.2: Performance and cost data for a tokamak optimised for Nb-Ti toroidal field and central solenoid coils with minimised capital cost (*) and upgraded magnets of different superconductors (with maximised net electricity yield). Also shown are the power values that would result, were a reduced H$_\text{98}$ = 1.0 to occur in practice.}
\begin{indented}\lineup
\item[]\begin{tabular}{@{}lllllllllllll} 
    \br
    &&&&&&&& \multicolumn{2}{c}{H$_\text{98}$ = 1.2} & \multicolumn{2}{c}{H$_\text{98}$ = 1.0}\\\cline{9-12}
    
    TF s.c.& CS s.c.&\makecell[l]{TF (CS)\\ peak field\\(\si{\tesla})}&\makecell[l]{Field on\\plasma\\(\si{\tesla})}&\makecell[l]{TF (CS)\\steel \%\\(\%)}&\makecell[l]{TF (CS)\\peak stress\\ (\si{\mega\pascal})}&\makecell[l]{Plasma\\current\\(M\si{\ampere})}&\makecell[l]{Capital\\cost\\(1990 M\$)}&\makecell[l]{Fusion\\power\\(M\si{\watt})}&\makecell[l]{Net.\\elec.\\(M\si{\watt})}&\makecell[l]{Fusion\\power\\(M\si{\watt})}&\makecell[l]{Net.\\elec.\\(M\si{\watt})}\\
    \br
   \makecell[l]{Commer.*\\Nb-Ti}&\makecell[l]{Commer.*\\Nb-Ti}&9.2 (8.7)&3.4&48.3 (81.6)&660 (660)&29.4&7730&2180&500&1500&270\\
    \mr
						
    \makecell[l]{Commer.\\Nb-Ti}&REBCO&9.2 (9.3)&3.4&48.3 (93.4)&660 (660)&29.6&7910&2220&510&1570&290\\
    									
    REBCO&REBCO&10.0 (11.9)&3.6&48.8 (92.2)&660 (660)&32.2&9790&2730&680&1730&340\\
    										
    \makecell[l]{Quat.\\Nb-Ti}&\makecell[l]{Quat.\\Nb-Ti}&9.4 (9.1)&3.4&47.8 (83.3)&660 (660)&30.1&9180&2290&540&1560&290\\
    										
    \makecell[l]{Quat.\\Nb-Ti}&REBCO&10.0 (9.4)&3.6&48.8 (92.2)&660 (660)&32.2&9600&2730&670&1730&340\\
    										
    \br
    \end{tabular}
\end{indented}
\end{table}
\end{landscape}

\begin{landscape}
\begin{table}
\caption{\label{CopperPilot}Performance and cost of a steady-state H$_\text{98}$ = 1.6,  100 \si{\mega\watt}$_\text{e}$ reactor with copper TF, CS and PF coils (with minimised capital cost). Based on STPP \cite{wilson04}.}
\begin{indented}\lineup
\item[]\begin{tabular}{@{}llllllllllllll} 
    \br
    &&&&&&&&&&\multicolumn{2}{c}{H$_\text{98}$ = 1.6}& \multicolumn{2}{c}{H$_\text{98}$ = 1.0}\\\cline{11-14}
    $R_\text{major}$&$A$&$\kappa$&$\delta$&\makecell[l]{Field on\\TF coil\\(\si{\tesla})}&\makecell[l]{Field on\\plasma\\(\si{\tesla})}&\makecell[l]{TF steel\\fraction\\(\%)}&\makecell[l]{Plasma\\current\\(M\si{\ampere})}&\makecell[l]{Capital\\cost\\(1990 M\$)}&~~&\makecell[l]{Fusion\\power\\(M\si{\watt})}&\makecell[l]{Net.\\elec.\\(M\si{\watt})}&\makecell[l]{Fusion\\power\\(M\si{\watt})}&\makecell[l]{Net.\\elec.\\(M\si{\watt})}\\
    \mr
    3.42&1.40&3.20&	0.55&	7.56&	2.50&5&	19.4&	5200&~~&		1670&	100&530&-370\\
    \br
    \end{tabular}
\end{indented}
\end{table}
\end{landscape}

\clearpage

\end{document}